\documentclass[prx,twocolumn,showpacs,preprintnumbers,amsmath,amssymb]{revtex4-1}
\usepackage{graphicx}
\usepackage{soul}

\begin{document}
\title{ Classification of magnons in Rotated Ferromagnetic Heisenberg model and their competing  responses in transverse fields  }
\author{ Fadi Sun$^{1,2}$, Jinwu Ye$^{1,3,4}$, and Wu-Ming Liu$^{2}$  }
\affiliation{
$^{1}$Department of Physics and Astronomy,
Mississippi State University, MS, 39762, USA \\
$^{2}$Beijing National Laboratory for Condensed Matter Physics,
Institute of Physics, Chinese Academy of Sciences,
Beijing 100190, China   \\
$^{3}$Key Laboratory of Terahertz Optoelectronics, Ministry of Education, Department of Physics, Capital Normal University,
Beijing, 100048, China
$^{4}$  Kavli Institute of Theoretical Physics, University of California, Santa Barbara, Santa Barbara, CA 93106  }

\date{\today }

\begin{abstract}
% In a previous work, we discussed the Rotated Ferromagnetic Heisenberg model (RFHM) which can be realized with
% spinor bosons loaded in optical lattices in the presence of spin-orbit coupling (SOC).
Competing orders is a general concept to describe various quantum phases and transitions in various materials.
One efficient way to investigate competing orders is to first classify different class of excitations in a given quantum phase,
then study their competing responses
under various external probes. This strategy may not only lead to deep understanding of the quantum phase itself, but also
its deep connections to various other quantum phases nearby. We implement this approach by studying
the Rotated Ferromagnetic Heisenberg model (RFHM) in two different transverse fields $h_x$ and $h_z$
which can be intuitively visualized as studying  spin-orbit couplings (SOC) effects in 2d Ising or anisotropic XY model in a transverse field.
%The RFHM is a new class of quantum spin model to study magnetisms in cold atom systems or materials with strong .
At a special SOC class, it was known that the RFHM at a zero field¡¡owns an exact ground state called Y-x state. It supports
non only the commensurate C-C$_0$ and C-C$_{\pi} $ magnons, but also the in-commensurate C-IC magnons.
These magnons are non-relativistic, not contained in the exact ground state, so need to be thermally excited.
Their dramatic response under the longitudinal $ h_y $ field was recently worked out by the authors.
Here we find they respond very differently under the two transverse fields.
Any $h_x$ ($h_z$) changes the collinear Y-x state to a canted co-planar YX-x (YZ-x) state which suffers quantum fluctuations.
The C-C$_0$, C-C$_{\pi} $ and C-IC magnons sneak into the quantum ground state, become relativistic and play leading roles even at $ T=0 $.
We map out the boundaries among the  C-C$_0$, C-C$_{\pi} $ and C-IC magnons,
especially the detailed evolution of the C-IC magnons inside the canted phases.
As $h_x$ ($h_z$) increases further, the  C-C$_0$ magnons always win the competition and
emerge as the seeds to drive a transition from the YX-x (YZ-x) to the X-FM ( Z-FM )
which is shown to be in the 3d Ising universality class.
%In sharp contrast to the Y-x state which is an exact ground state which contains no information on
%The YX-x ( or YZ-x ) state is just a classical state which receive strong quantum corrections.
We evaluate their contributions to magnetization, specific heat, uniform and staggered longitudinal susceptibilities
spin correlation functions which can be measured by various established experimental techniques.
Some analogies with quantum fluctuations generated  multiple vortices and multiple landscapes in quantum spin glass are mentioned.
The implications to cold atom systems and materials with SOC are briefly discussed.
% We determine the orbital and spin structures and
% the novel excitation spectra of  all the quantum phases. We also explore new universality classes of
% quantum phase transitions between( or multi-critical point among ) these phases.
%  with the dynamic exponents $ z=1, z=2 $ and the anisotropic one $ ( z_x=1, z_y=3 ) $
% and finite temperature transitions above all these phases.
%  The SOC in a lattice provides to a new platform to generate whole new classes of quantum phases and phase transitions.
% Possible implications to condensed matter systems with or without SOC are given.
\end{abstract}

\maketitle

\section{Introduction}
A fundamental problem in any branch of physics is on the nature of the ground state.
In strongly correlated electron systems, competing orders is a general concept to describe
various quantum phases and transitions in various materials.
In this work, we focus on competing excitations in a given quantum phase which could lead to a natural explanation of some competing orders.
It is known a given quantum phase can still support different kinds of excitations with their own characteristics.
These different classes of excitations are generated by quantum fluctuations inherent in the quantum ground state, so
are intrinsic objects embedded in the ground state itself. Under various external probes, these excitations
compete to emerge to drive the instability of the system into various other quantum phases through different universality classes
of quantum phase transitions. So classifying different classes of excitations of a given phase and
investigating their behaviors under various external probes could lead to deep understandings not only on the nature of the ground state itself, but
also its deep connections to various other quantum phases.

%So classifying different classes of excitations of a given phase not only lead to deep understanding
%of the nature of the ground itself, also lead to its deep connections to various other quantum phases.

A quantum phase is characterized by its symmetry breaking and excitation spectrum \cite{scaling,aue,wenbook,sachdev,frad}.
For quantum spin or bosonic systems \cite{scaling,sachdev}, gapless excitations indicate long-range correlations encoded in the quantum phase.
External probes could open a positive gap to the excitation or induce a " negative" gap which indicates a quantum phase transition to another phase.
While gapped excitations \cite{rh,rhh} indicate short-ranged fluctuations encoded in the phase.
The external probes such as magnetic fields, pressures, electric fields, etc may drive these gapped excitations near to a QCP,
close their gaps and lead to their condensations into a new phase through a quantum phase transition.
For fermionic systems \cite{blqh,tri,tqpt,ssdw,pairing,rafhm}, the quantum phase supports both
fermionic excitation and collective bosonic excitations. The two sectors may compete to lead to various other quantum phases under various external
probes. Due to the absence of any symmetry breaking, a topological phase ( such as quantum Hall state, spin liquids )  \cite{wenbook,frad,kitpconf}
is characterized by its topological orders and associated fractionalized excitations. The gap closings of these
fractionalized excitations could lead to another topological phase through a Topological phase transition.
In this work, we only focus on quantum phases without topological orders and with only different kinds of bosonic excitations.

So far, the classifications of the quantum ground states in terms of
symmetry breaking and possible resulting gapless Goldstone modes are well studied \cite{scaling,aue,wenbook,sachdev,frad}.
However, much less attentions have been paid to classify gapped excitations in a quantum phase.
In a previous work \cite{rh}, the authors studied the Rotated Ferromagnetic Heisenberg model (RFHM) which is a new
class of quantum spin model to describe quantum magnetisms in some cold atom systems or materials with strong spin-orbit coupling (SOC)
\cite{cwu,classdm1,classdm2}.
For a specific SOC class, we identify a new spin-orbital entangled commensurate ground state: the Y-x state.
It supports 3 kinds of magnons: commensurate magnons such as $ C-C_0, C-C_{\pi} $  and also a new gapped elementary excitation:
in-commensurate magnon ( C-IC ) with its two gap minima continuously
tuned by the SOC strength. They are gapped bosonic excitations taking non-relativistic dispersion with anisotropic
effective mass $ m_x, m_y $.
However, the Y-x ground state is an exact quantum ground state with no quantum fluctuations.
So the $ C-C_0, C-C_{\pi}, C-IC $ magnons in the RFHM are extrinsic,
     not embedded in the ground state due to the absence of quantum fluctuations.
     They need to be excited by thermal fluctuations
     Their parameters such as the minimum positions $ ( 0, \pm k^{0}_y ) $, gap $ \Delta $, masses $ m_x, m_y $ can
     only be measured by various characteristics of the transverse structure factor at a finite $ T $:
     it is a Gaussian shape, peaked at  $ ( 0, \pm k^{0}_y ) $ with an exponentially suppressed amplitude $ e ^{-\Delta/T} $,
     with a temperature dependent width  $ \sigma_x=\sqrt{ m_x(\beta) T } $.

The existence of the C-IC above a commensurate phase is the most striking feature of the RFHM. They indicate the short-ranged
in-commensurate order embedded in a long-range ordered commensurate ground state. They are
the seeds driving possible transitions from commensurate to another commensurate phase with different
spin-orbital structure or to an In-commensurate phase. An important question to ask is how to drag out these C-IC magnons,
closing their gap and drive into new quantum phases
through the condensation of these magnons ? In a recent work \cite{rhh}, the authors showed that applying a
uniform longitudinal Zeeman field $ h_y $ could do the job very well:  the $ C-C_0, C-C_{\pi}, C-IC $ magnons
compete to emerge under its effects to drive quantum phase transitions.
It turns out that the C-IC always win the competition.
Indeed, the Zeeman field drives the condensation of the C-IC, therefore, the Y-x state to
non-coplanar incommensurate Skyrmion (IC-SkX) crystal phases through new universality class of quantum phase transitions.
At intermediate fields, the IC-SKX will turn into two co-planar canted states at two dual related SOC parameters respectively.
At high fields, it will become a spin polarized state with quantum fluctuations.

Because the SOC breaks the spin rotation symmetry,
so the two transverse fields $ h_x, h_z $ may play quite different roles in dragging out the C-IC magnons, therefore
drive into completely different phases and phase transitions than the longitudinal field $ h_y $ studied in \cite{rhh}.
Following \cite{rh,rhh}, in this work, we also focus along the solvable line $ (\alpha=\pi/2, \beta) $ of the RFHM in a transverse field Eqn.\ref{hhx}
and Eqn.\ref{hhz}. Away from $ \alpha=\pi/2 $ will be briefly mentioned  in the conclusion section and be presented in a separate publication.
The two models can be considered as incorporating possible dramatic effects of SOC on well studied
2d Ising, anisotropic ( or isotropic ) quantum XY model in a transverse field.
Note that the $ h_y $ field keeps the hidden $ U(1) $ symmetry
$ \sum_{i} (-1)^{x} S^y_i $ of the Hamiltonian at the zero field, but $h_x$, $h_z$ breaks it.
This fact alone may lead to dramatic different competition among the magnons
when they are subject to the longitudinal $ h_y $ or the two transverse fields $h_x$ and $h_z$ \cite{sachdev}.
Indeed, in the longitudinal field $ h_y $,  there is a Mirror transformation \cite{rhh} relating $ ( \beta, h_y ) $ to  $ ( \pi/2-\beta, h_y ) $.
So $ \beta=\pi/4 $ enjoys the Mirror symmetry. However, because $ h_x $ and $ h_z $ explicitly breaks the $ U(1) $ symmetry,
so the mirror transformation does not work anymore in $ h_x $ and $ h_z $ case.
However, in the $ h_x $ case, we will still able to find a ¡°generalized¡± mirror transformation
to characterize systematically the competitions among the magnons. Unfortunately, there is even no such ¡°generalized¡± mirror transformation
in the $ h_z $ case, so the competitions are more intricate in the $ h_z $ field.

In the $ h_x $ field, any infinitesimal $ h_x $ will change the Y-x state into a canted YX-x  state.
In sharp contrast to the Y-x state which is an exact ground state free of quantum fluctuations.
The YX-x state suffers quantum fluctuations.
So at $ T=0$, these magnons are quantum fluctuations generated, sneak into the YX-x state and become
important components embedded inside the quantum ground state.
They stand for quantum fluctuations with intrinsic wavelength and frequency,
so can be detected by spin structure factor even at $ T=0 $. We also evaluate their contributions
to magnetization, specific heat, uniform and staggered longitudinal susceptibilities at a finite temperature.
Using the ¡°generalized¡± mirror transformation, we map out the boundaries of the commensurate magnons C-C$_0$, C-C$_{\pi}$ and the in-commensurate magnons C-IC
inside the YX-x canted phase. As $ h_x $ increases, the C-C$_0$ magnons emerge and drives the quantum phase transition
at a critical field $ h_{cx}(\beta) $ from the YX-x phase to the X-FM phase.
By identifying a suitable order parameter, performing spin wave expansion and symmetry analysis, we find it is in 3d Ising the universality class.
Due to the enlarged U(1) symmetry, the transition at the Abelian $ \beta=\pi/2 $ point is driven by the simultaneous condensations of
the C-C$_0$ and C-C$_{\pi}$ magnons and is in the universality class of 3d $ XY $ model.
%     We determine the universality class of the transition which is controlled by a fixed point on the phase boundary.
%     So the transition at any non-abelain point is different that at the
%     Abelian end point $ \beta=\pi/2 $. This fact is rooted at $ h_x $ breaks the  $ U(1) $ symmetry $ \sum_{i} (-1)^{x} S^y_i $ away from the Abelian point.

In the $ h_z $ field, by applying the hidden $ U(1) $ symmetry operator at the zero field case, we show that the $ h_z $ case can be mapped to RFHM
in a staggered  $ h_x $ field along the $ x $ direction.
We also work out the corresponding phase diagram in this case.
We show that any infinitesimal $ h_z $ will change the Y-x state into a canted YZ-x  state,
then a phase transition into the Z-FM state at a critical field $ h_{cz}(\beta) $ which
is shown to be also in the universality class of 3d Ising model. We find the transition from the YZ-x canted phase to the Z-FM
is always driven by the condensations of the C-C$_0$ magnons.
Unfortunately, the ¡°generalized¡± mirror transformation used in the $ h_x $ case does not work  in the $ h_z $ case anymore,
this fact makes the landscapes of the C-IC magnons much more complicated in the $ h_z $ case than the $ h_x $ case.
Even so, we are still able to map out the competing boundaries and detailed structures of the C-C$_0$, C-C$_{\pi}$
and C-IC magnons inside the YZ-x canted phase.
%work out most details of the structure of the C-IC magnons which indicate the intrinsic quantum fluctuation patterns embedded in the ground state.
Due to the enlarged U(1) symmetry, the transition at the two Abelian points $ \beta=0 $ and $ \beta= \pi/2 $ point
is driven by the condensation of $ C-C_0 $ and the simultaneous condensations of
the C-C$_0$ and C-C$_{\pi}$ magnons respectively and is in the universality class of 3d $ XY $ model.
In principle, all the thermodynamic quantities such as the magnetization, specific heat, uniform and staggered susceptibilities in the YZ-x canted phase, the Z-FM and their quantum critical scalings can be calculated.

In view of recent remarkable experimental realization of 2d Rashba or Dresselhaus SOC
or any of their linear combinations in Fermi gas or spinor BEC \cite{expk40,2dsocbec}.
The two models Eqn.\ref{hhx} and Eqn.\ref{hhz} can be realized in near future experiments.
The results achieved in this work can be  detected by various techniques such as  specific heat  \cite{heat1,heat2}, In situ measurement \cite{dosexp} and light or atom Bragg spectroscopy \cite{lightatom1,lightatom2} respectively. They may also shed some lights to study magnetic orderings
in some  strongly correlated SOC materials \cite{kitpconf,rhh} with $ h_x, h_z $ play the roles of crystal fields.

\section{ Transverse field $ h_x $ }

The Rotated Ferromagnetic Heisenberg model (RFHM) \cite{rh}  in a transverse field along the $S_x$ direction
is described by:
\begin{align}
    \mathcal{H}
    \!=\!-\!J\!\sum_i
    [\mathbf{S}_i R_x( 2 \alpha ) \mathbf{S}_{i+x}
    \!\!+\!\mathbf{S}_i R_y(2\beta) \mathbf{S}_{i+y}]
    \!-\!H_x\!\!\sum_i \!S_i^x
\label{hhx}
\end{align}
where $J>0$ is the ferromagnetic interaction
and the sum is over a unit cell $i$ in a square lattice,
the $R_x(2\alpha)$, $R_y(2\beta)$ are two SO(3)
rotation matrices around $\hat{x}$, $\hat{y}$ spin axis
by angle $2\alpha$, $2\beta$ putting on the two bonds
$\hat{x}$, $\hat{y} $ respectively.
Following the previous works \cite{rh,rhh}, we also focus along the solvable line $ (\alpha=\pi/2, \beta) $.
At the zero field case $ H_x=0 $, the ground state is the $ Y-x $ state shown in the horizontal axis in Fig.\ref{phasehx}.

As shown in \cite{rh}, at  $ H_x=0 $, the Hamiltonian has the Time reversal ( $ {\cal T } $ )  symmetry, translational symmetry and 3 spin-orbital coupled
$ Z_2 $ symmetries $ {\cal P}_x, {\cal P}_y, {\cal P}_z $, especially a spin-orbital coupled $ U(1) $ symmetry.
However, any $ H_x $ will break all these symmetries except the $ {\cal P}_x $ symmetry:
$ S^{x} \rightarrow S^{x}, k_y \rightarrow - k_y, S^{y} \rightarrow - S^{y},  S^{z} \rightarrow - S^{z} $and the translational symmetry.
It also keeps the combined  $   {\cal T } {\cal P}_y $ and $ {\cal T } {\cal P}_z $ symmetries.

\subsection{ X-FM state and excitations in the strong field  }

\begin{figure}[!htb]
    \includegraphics[width=\linewidth]{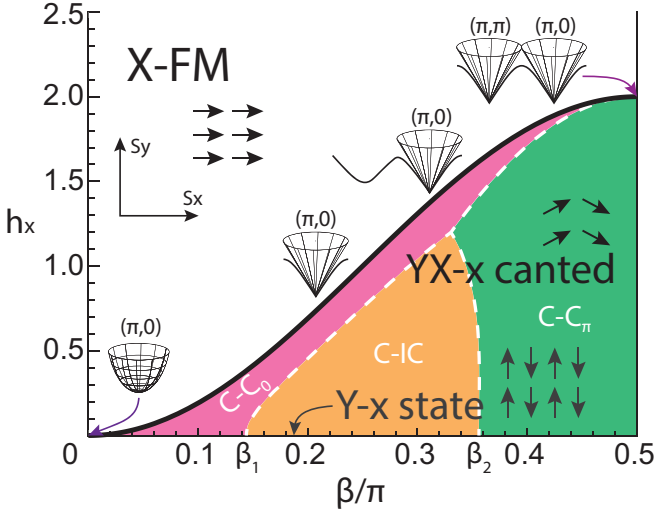}
\caption{  ( Color online )  Phase diagram of the RFHM in a transverse field $h_x$.  Boundaries between C-C$_0$,C-C$_\pi$,C-IC
	     are indicated by white dashed lines.
         At $ h_x=0 $, the boundary between  C-C$_0$ and C-IC is $\beta_1$ and  that between C-C$_\pi$ and C-IC is $\beta_2$,
	     as shown in  Ref.\cite{rh}. They are non-relativistic.
         At $h_x=0$, the ground state is the Y-x state which is exact without any quantum fluctuations.
         So it contains no information  of C-C$_0$, C-C$_\pi$ and C-IC.
         Any $ h_x > 0$ will transfer it into the YX-x  canted state which suffers quantum fluctuations.
         The C-C$_0$, C-C$_\pi$ and C-IC  sneak into the YX-x ground state and become relativistic.
         At the critical field $ h_c(\beta) $, the $ YX-x $ canted state undergoes  a quantum phase transition to the X-FM state.
         It is  in the 3d Ising university class and always driven by the condensation of $ C-C_0 $ magnons.
         The $ C-IC $ and  C-C$_\pi$ magnons  always lose to the $ C-C_0 $ magnons before hitting the phase boundary.
         However, at the Abelian point $ \beta=\pi/2 $, $ C-C_0 $ and  C-C$_\pi$ magnons condense simultaneously and lead to a
         3d XY class transition to the X-FM.
         The fine landscape of the  C-IC regime is given in Fig.\ref{finehx} and also discussed in details in Appendix C. }
\label{phasehx}
\end{figure}

To map out the phases of Eqn.\ref{hhx}, it is instructive to start from the high field limit  $H_x\gg J$.
In this limit, the system is in X-FM phase
with spin  fully ( classically ) polarized to $S_x$ direction Fig.\ref{phasehx}.
Obviously, the X-FM keeps all the symmetry of the Hamiltonian.

 Under the global spin rotation $(S_i^x,S_i^y,S_i^z)\to(S_i^z,S_i^y,-S_i^x)$, Hamiltonian becomes
\begin{align}
    \mathcal{H}
    =-J\sum_i
    [\mathbf{S}_i R_z(\pi) \mathbf{S}_{i+x}
    +\mathbf{S}_i R_y(2\beta) \mathbf{S}_{i+y}]
    -H_x\sum_i S_i^z
\label{hhxz}
\end{align}
Introducing the Holstein-Primakoff (HP) bosons \cite{rh,rhh},
$S^+=\sqrt{2S-a^\dagger a}a$,
$S^-=a^\dagger\sqrt{2S-a^\dagger a}$,
$S^z=S-a^\dagger a$, to the linear spin wave ( LSW ) order at $ 1/S $, we map the Hamiltonian Eqn.\ref{hhxz} to:
\begin{align}
    \mathcal{H}
    \!=\!E_0\!+\!2JS
    \sum_k&\big[
    (h_x\!+\!\cos k_x\!+\!\cos^2\!\beta(2\!-\!\cos k_y))
    a_k^\dagger a_k  \nonumber\\
    &+\sin^2\beta\cos k_y
    (a_ka_{-k}\!+\!a_k^\dagger a_{-k}^\dagger)/2
    \big]
\label{hxlsw}
\end{align}
where the classical ground state energy $E_0=-2JNS^2\cos^2\beta-H_xNS$
and we have introduced the dimensionless field $ h_x=H_x/(2JS) $.
Now the Hamiltonian can be diagonalized by a Bogoliubov transformation
\begin{align}
    \mathcal{H}
    \!=\!E'_0\!+\!4JS
    \sum_k \omega_k\alpha_k^\dagger\alpha_k
\end{align}
where the ground-state energy at the order of $1/S $ is
$E'_0=E_0-2JS\sum_k \omega_k$ and the energy spectrum is:
\begin{align}
    \omega_k\!=\!\!\sqrt{\!
    [h_x\!+\!\cos\! k_x\!+\!\cos^2\!\!\beta(2\!-\!\cos k_y)]^2
    \!-\!\sin^4\!\!\beta\cos^2\!k_y}
\label{spectrumhhx}
\end{align}
where, for $0<\beta<\pi/2$, one can identify that there is a unique minimum located at $k^0=(k_x,k_y)=(\pi,0)$
with the energy gap:
\begin{align}
    \Delta_{\pi}=\omega_{k=k^0}=
    \sqrt{h_x(h_x-1+\cos2\beta)}
\end{align}

   The gap vanishing condition leads to the critical field strength $ h_c $:
\begin{align}
    h_{cx}(\beta)=1-\cos2\beta=2\sin^2\beta
\label{hcx}
\end{align}
  which is shown in Fig.\ref{phasehx}.
  The gap vanishing at $k^0=(k_x,k_y)=(\pi,0)$ indicate a quantum phase transition into a spin-orbital correlated state with orbital order
  $ (\pi,0)$. It was known that at $ h_x=0 $, the ground state $ Y-x $ state also has the $ ( \pi,0) $ orbital order.
  That indicates that there is only one phase transition and the state below $ h_{cx} $ could be just the $ YX-x $ state with a canted angle.
  As to be shown in the next subsection, we show that it is indeed the $ YX-x $ state with the orbital order $ (\pi,0) $.
  So near the QPT, $ \Delta_{\pi} \sim (h_x - h_{cx} )^{1/2} $.

  From Eqn.\ref{spectrumhhx}, we find the excitation spectrum  around the minimum takes the relativistic form
\begin{align}
    \omega_q=\sqrt{\Delta^2_{\pi}+v_x^2 q_x^2+v_y^2 q_y^2}
    ,\quad
    k=k^0+q
\label{relmag1}
\end{align}
where
\begin{align}
    v_x^2&=(2h_x-1+\cos2\beta)/2     \nonumber  \\
    v_y^2&=[h_x+\cos2\beta(h_x-1+\cos2\beta)]/2
\label{twoxfm}
\end{align}

At $ h=h_{cx} $, the critical velocities are
$v_{x,c}^2=v_{y,c}^2=\sin^2\beta$. As long as $\beta>0$, we obtain a non-zero critical velocity,
which indicate a relativistic critical behavior with the dynamic exponent $ z=1 $.

At the two Abelian points $\beta=0,\pi/2$ the system has SU(2) symmetry in the rotated basis $ \tilde{SU}(2) $
with $\vec{\tilde{S}}_{i} = R(\hat{x},\pi n_1) \vec{S}_{i}$
and $ \tilde{\tilde{SU}}(2) $ with $ \vec{\tilde{S}}_{i} = R(\hat{x},\pi n_1)  R(\hat{y},\pi n_2) \vec{S}_{i} $ respectively.
So Eqn.\ref{hhx} can be mapped to a FM Heisenberg model in $ -h_x \sum_i \tilde{S}^x_{i} $ and $ -h_x \sum_i (-1)^{i_y} \tilde{\tilde{S}}^x_{i} $
( see Eqn.\ref{ttsu2} ) respectively.
So at $ \beta=0 $, any $ h_x $ will pick up the $ X-FM $ phase as the exact ground state. At $ \beta=\pi/2 $, taking the result from \cite{rhh},
any $ h_x $ will lead to a spin-flop transition resulting into a $ U(1) $ symmetry breaking canted phase with one Goldstone mode $ \phi $.
Then there is another transition to the X-FM at a finite $ h_c=2 $. These results at the two Abelian points are consistent with the general result Eqn.\ref{hcx}
and shown in Fig.\ref{phasehx}

At $\beta=0, h_x=0$, transferring back to the original basis, the Hamiltonian Eq.\ref{hhx} has the SU(2) symmetry
generated by $\sum_i S_i^x, \sum_i (-1)^{i_x}S_i^y, \sum_i (-1)^{i_x}S_i^z$.
At any $h_x>0$, only $\sum_i S_i^x$ remains as a conserved quantity.
Obviously, the X-FM state keeps all symmetries of the Hamiltonian.
Acting the conserved quantity $ e^{i \phi \sum_i S_i^x } $ on the minimum $(\pi,0)$,
changes nothing. So we conclude that at $\beta=0, h_x>0$,
the system has only one minima located at $(\pi,0)$ as shown in Fig.\ref{phasehx}.

Similarly, at $\beta=\pi/2, h_x=0$,  transferring back to the original basis, the Hamiltonian Eq.\ref{hhx} has the SU(2) symmetry
generated by $\sum_i(-1)^{i_y} S_i^x, \sum_i (-1)^{i_x}S_i^y, \sum_i (-1)^{i_x+i_y}S_i^z$.
At any $h_x>0$, only $\sum_i (-1)^{i_y} S_i^x$ remains as a conserved quantity.
Acting the conserved quantity $ e^{i \sum_i (-1)^{i_y} S_i^x } $ on the minimum $(\pi,0)$,
will generate another minimum $(\pi,\pi)$, we conclude that at
$\beta=\pi, h_x>0$, the system has two minima \cite{rhh} located at $(\pi,0)$ and $(\pi,\pi)$ as shown in Fig.\ref{phasehx}.

\subsection{ YX-x canted state below $h_{cx}$.  }

{\sl 1. Classical YX-x canted phase at $h<h_{cx}$ }

When $\beta=\pi/2$, in the  $ \tilde{\tilde{SU}}(2) $ basis $\tilde{\mathbf{S}}_i=R_x(i_x\pi)R_y(i_y\pi)\mathbf{S}_i$,
the Hamiltonian Eq.\ref{hhx} takes the form:
\begin{align}
    \mathcal{H}=-J\sum_{\langle ij\rangle}
    \tilde{\mathbf{S}}_i\cdot\tilde{\mathbf{S}}_j
    -H_x\sum_i (-1)^{i_y}\tilde{S}_i^x
\label{ttsu2}
\end{align}
  When $0<H_x<H_{xc}$, the classical state takes the form:
\begin{align}
    \tilde{\mathbf{S}}_i=S
    \big((-1)^{i_y}\cos\theta,\sin\theta\cos\phi,\sin\theta\sin\phi\big)
\label{statex1}
\end{align}
 where $ \phi $ is the Goldstone mode.

Reverting back to the original basis, we obtain the classical state in the original basis:
\begin{align}
    \mathbf{S}_i\!=\!S
    \big(\cos\theta,
	 (-1)^{i_x}\!\sin\theta\cos\phi,
	 (-1)^{i_x\!+i_y}\!\sin\theta\sin\phi
    \big)
\label{statex2}
\end{align}

Note that although we obtained Eq.\eqref{statex1} and Eq.\eqref{statex2}
at $\beta=\pi/2$, the same ansatz hold for $ 0<\beta<\pi/2 $ whose classical ground energy is:
\begin{align}
    E_{\rm c}
    \!=\!-2NJS^2(1\!+\!h_x\cos\theta\!-\!\sin^2\!\beta\cos^2\!\theta
     \!-\!\cos^2\!\beta\sin^2\!\theta\sin^2\!\phi)
\label{classex}
\end{align}

 It is easy to see that any deviation from $ \beta=\pi/2 $ explicitly breaks the $ U(1) $ symmetry at $ \beta=\pi/2 $,
 so picks up $\phi=0$ and leads to the YX-x canted state:
\begin{align}
    \mathbf{S}_i=S
    \big( \cos\theta,(-1)^{i_x}\!\sin\theta,0 \big)
\label{yxx}
\end{align}
   which indeed has the $ (\pi,0 ) $ order as indicated from the magnon condensations from the X-FM studied in the subsection A.

  Substituting $\phi=0$ in Eq.\eqref{classex} leads to the classical ground-state energy
\begin{align}
    E_{\rm c}=
    -2NJS^2(1+h_x\cos\theta-\sin^2\beta\cos^2\theta)
\label{classex0}
\end{align}
   whose minimization leads to the canted angle:
\begin{align}
    \cos\theta=\frac{h_x}{2\sin^2\beta}<1,
    \quad
    \textmd{ when $h_x<h_{cx}$}
\label{canthx}
\end{align}
  which always has a solution as long as $ h_x<h_{xc}$.

 Only when $ h=h_{cx}, \theta=0 $, it becomes the X-FM phase.
 The fact that we achieved the same critical field $ h_{cx} $ from
 the X-FM state Eqn.\ref{hcx} above it and the YX-x state Eqn.\ref{canthx} below it indicate that there is only one phase transition
 with the critical field   $ h=h_{cx} $ shown in Fig.\ref{phasehx}.
 Note that from above $ h > h_{cx} $, we achieved it by the LSW at the order of $ 1/S $.
 while, from below $ h > h_{cx} $, we achieved it just by the classical ground state energy minimization at $ S=\infty $.
 
 In sharp contrast, in the $ h_y $ case \cite{rhh}, there are two critical fields $ h_{c1} < h_{c2} $, there is an 
 intermediate  IC-SKX phase between the two critical fields.

{\sl 2. Spin wave analysis in the YX-x Canted state }

Again performing the global spin rotation
$(S_i^x,S_i^y,S_i^z)\to(S_i^z,S_i^y,-S_i^x)$,
then applying the spin rotation $R_x(\theta)$ for the $A$-sublattice
and $R_x(-\theta)$ for the $B$-sublattice lead to:
\begin{align}
    \mathcal{H}=
    &-J\sum_{i\in A}
    [S_i R_z(\pi) S_{i+x}
    +S_iR_x(\theta)R_y(2\beta)R_x(-\theta)S_{i+y}] \nonumber\\
    &-J\sum_{i\in B}
    [S_i R_z(\pi) S_{i+x}
    +S_iR_x(-\theta)R_y(2\beta)R_x(\theta)S_{i+y}] \nonumber\\
    &-H_x\sum_{i\in A}
    [\sin\theta S_i^y+\cos\theta S_i^z] \nonumber\\
    &-H_x\sum_{i\in B}
    [-\sin\theta S_i^y+\cos\theta S_i^z]
\label{Hyxx}
\end{align}

Introducing the Holstein-Primakoff (HP) bosons
$S^+=\sqrt{2S-a^\dagger a}a$,
$S^-=a^\dagger\sqrt{2S-a^\dagger a}$,
$S^z=S-a^\dagger a$ for sublattice $A$
and
$S^+=\sqrt{2S-b^\dagger b}b$,
$S^-=b^\dagger\sqrt{2S-b^\dagger b}$,
$S^z=S-b^\dagger b$ for sublattice $B$,
we map the Hamiltonian Eq.\ref{Hyxx} at the order $ 1/S $ to
\begin{align}
    \mathcal{H}=E_c
    +2JS\sum_k
    [(A_k\!+\!B_k)a_k^\dagger a_k\!+\!
     (A_k\!-\!B_k)b_k^\dagger b_k \nonumber\\
    +C_k (a_k^\dagger b_k+b_k^\dagger a_k)
    +D_k (a_k a_{-k}+b_kb_{-k}+h.c.)]
\label{Hyyx1s}
\end{align}
  where $E_c$ is the classical ground state energy Eq.\eqref{classex0} and
\begin{align}
	&A_k=2-(\cos^2\beta-\sin^2\beta\sin^2\theta)\cos k_y  \nonumber  \\
	&B_k=\sin2\beta\sin\theta\sin k_y    \nonumber \\
	&C_k=\cos k_x    \nonumber \\
	&D_k=\sin^2\beta \cos^2\theta\cos k_y
\label{abcdhx}
\end{align}

      The Hamiltonian Eqn.\ref{Hyyx1s} can be diagonalized by a Bogoliubov transformation
\begin{align}
    \mathcal{H}=E'_c+4JS\sum_k
    (\omega_k^-\alpha_k^\dagger\alpha_k
    +\omega_k^+\beta_k^\dagger\beta_k)
\end{align}
where $E'_c=E_c-2JS\sum_k(\omega_k^-+\omega_k^+)$  is the ground state energy up to the order of $1/S $
and the energy spectra are:
\begin{align}
	\omega_k^{\pm}\!=\!\!
	\sqrt{\!A_k^2\!+\!B_k^2\!+\!C_k^2\!-\!D_k^2\!\pm\!
	2\sqrt{\!A_k^2(B_k^2\!+\!C_k^2)\!-\!B_k^2D_k^2}}
\label{YXxen1s}
\end{align}
from which one can determine the minimum positions.

We found there are three regimes inside the YX-x Canted state:
C-C$_{0}$ regime, C-IC regime, and  C-C$_{\pi}$ regime which, at $ h_x=0 $, reduce to
the three regimes identified in \cite{rh}.
Among the three regimes, only C-C$_{0}$ regime sits just below the transition line $ h_{cx} $,
so the transition from the YX-x state to the X-FM is driven by the condensations of the C-C$_{0}$ magnons only.

Because the transition from YX-x Canted state to X-FM state is driven by the condensation of the C-C$_0$ magnons.
We find that just below the phase boundary, the C-C$_0$ magnons take also the relativistic form around $k_0=(0,0)$:
\begin{align}
    \omega_q=\sqrt{\Delta^2_0+v_x^2 q_x^2+v_y^2 q_y^2},
    \quad
    k=q+k_0
\label{relmag2}
\end{align}
where
\begin{align}
    \Delta_0&=\sqrt{
	(1-\cos2\beta)
	(1-\cos2\beta-\frac{h_x^2}{2\sin^2\beta})},   \nonumber \\
    v_x^2&=2\sin^2\beta-(\frac{h_x}{2\sin\beta})^2,  \nonumber \\
    v_y^2&=\Big[2\sin^2\beta-(\frac{h_x}{2\sin\beta})^2\Big]\nonumber\\
         &\times\Big[\cos2\beta\!-\!\sin^22\beta
		\!+\!\frac{h_x^2}{4\sin^2\beta}
		\!+\!\frac{h_x^2\cos^2\beta}{\sin^2\beta}
	   \Big]\nonumber\\
	   &+\Big(\frac{h_x}{2\sin\beta}\Big)^4\!
	   \Big[
		1\!+\!\frac{\sin^22\beta(1\!-\!\frac{h_x^2}{4\sin^4\beta})}
		       {2\!-\!\cos2\beta\!-\!(\frac{h_x}{2\sin\beta})^2}
	   \Big]
\end{align}

At $ h=h_{cx} $, the critical velocities are $v_{x,c}^2=v_{y,c}^2=\sin^2\beta$ which are the same as those achieved from X-FM from above
the $ h_{cx} $ in Eqn.\ref{twoxfm}. Near $ h_{cx} $, $ \Delta \sim ( h_{cx}-  h )^{1/2} $.
Now we can check the consistence of the orbital orders on both sides of $ h_{cx} $.
The YX-x state has the orbital order $ (\pi,0) $, the $ C-C_0 $ has the orbital order $ (0,0)=(\pi,0) $ in the RBZ.
So its condensation on the top of  YX-x could lead to the two orbital orders either $  (\pi,0)+(0,0)=(\pi,0) $
or $  (\pi,0)+(\pi,0)=(0,0) $ in the EBZ. The $ (0,0) $ order  is nothing but that of the X-FM in Fig.\ref{phasehx}

 The competition between C-C$_0$ and C-C$_{\pi}$ gives
the boundary between C-C$_0$ and C-C$_{\pi}$  where they become degenerate:
\begin{equation}
   h_{0\pi}=2\sin\beta\sqrt{-\cos2\beta} < h_{cx}
\label{minihx3}
\end{equation}
  where $ \beta^{*} \sim 0.330482\pi  < \beta < \pi/2 $.

The competition  between C-C$_0$ and C-IC is given by the condition:
$ \frac{\partial^2 \omega_k^-}{\partial k_y^2}|_{k=(0,0)}=0$.
That between C-C$_\pi$ and C-IC is given by the condition:
$ \frac{\partial^2 \omega_k^-}{\partial k_y^2}|_{k=(0,\pi)}=0$.
We find that the three boundaries (dashed lines) in Fig.\ref{phasehx}
meet at the same point $(\beta^{*}=0.330482\pi,h^{*}_x=1.19921)$.
The fine structure near this point is shown in Fig.\ref{finehx}

\subsection{ Evolution of the $ C-IC $ magnons inside the $ C-IC $ regime in Fig.\ref{phasehx}:
generalized Mirror symmetry about the contour $ k^{0}_y= \pm \pi/2 $.  }

\begin{figure}[!htb]
	%\centering	\includegraphics[scale=0.6]{phase_n}\quad
\includegraphics[scale=0.52]{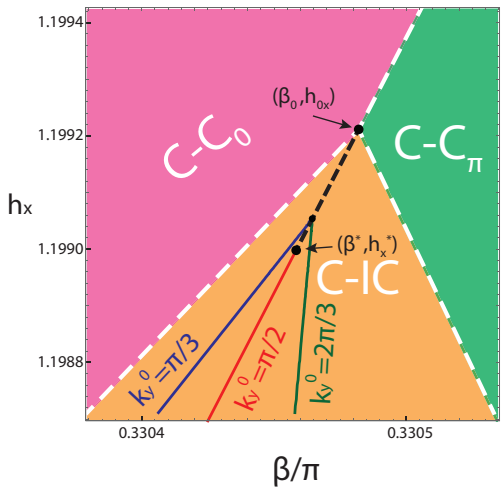}
\caption{  ( Color online ) The evolution of the C-IC magnons in Fig.\ref{phasehx} zoomed ($\times10^4$)
    around $ ( \beta_0 = 0.330458\pi, h_{0x}=1.19899 ) $ and $(\beta^{*}=0.330482\pi,h^{*}_x=1.19921)$.
    There is a generalized mirror symmetry around $ k^{0}_{y}=\pm \pi/2 $.
    The minimum at $ ( 0, \pm k^{0}_{y} ) $ and its mirror image at  $ ( 0, \pm ( \pi- k^{0}_{y}) ) $ symmetrically located on the two sides
    of  $ k^{0}_{y}=\pm \pi/2 $ must end in the regime $  \beta_0 < \beta < \beta^{*} $. The three segments of the contour
    line $ k^{0}_{y}=\pm \pi/2 $ are explained in the text. }
\label{finehx}
\end{figure}

In the longitudinal field $ h_y $ which keeps the spin-orbital coupled $ U(1) $ symmetry \cite{rhh},
there is a Mirror transformation relating $ ( \beta, h_y ) $ to  $ ( \pi/2-\beta, h_y ) $.
So $ \beta=\pi/4 $ enjoys the Mirror symmetry.
However, because $ h_x $ and $ h_z $ explicitly breaks the $ U(1) $ symmetry, so the mirror transformation does not
work anymore in $ h_x $ and $ h_z $ case.
Even so, it would be important to first understand the constant contour minimum  at $  k^{0}_{y}= \pm \pi/2 $.
In the $ h_x $ case, it seems there is a "generalized" Mirror transformation relating the minimum at
$ ( 0, \pm k^{0}_{y} ) $ to its associated mirror image at $ ( 0, \pm (\pi- k^{0}_{y}) ) $ as shown in Fig.\ref{finehx},
while the $  k^{0}_{y}= \pm \pi/2 $ is the self-dual line which starts at $ (\beta=\pi/4, h_x=0 ) $.
Unfortunately, in contrast to the $ h_y $ case, it is difficult to find the exact form of such a
"generalized" Mirror transformation in terms of $ (\beta, h_x ) $. Its form in term of the contour $  k^{0}_{y} $
would be enough to analyze the structure of the C-IC regime in Fig.\ref{finehx} at least to the order of $ 1/S $.
However, as to be shown in the next section, there is no "generalized" Mirror symmetry in the $ h_z $ case.

As shown in the Appendix B, the minimum contour $k_y^0= \pm \pi/2$ can be determined by the equation
\begin{align}
	h_{\pi/2}(\beta)=2\sin\beta\sqrt{-\cos2\beta}
\label{minihx1}
\end{align}
  where $ \quad 0.25\pi<\beta<\beta_0\approx0.330458\pi $.

If comparing Eqn.\ref{minihx1} with the C-C$_0$/C-C$_\pi$ boundary Eqn.\ref{minihx3},
we will find out they have the same form but different domains.
In fact, one can extend Eq.\ref{minihx1} to the whole domain $0.25\pi<\beta<0.50\pi$ where we have two special $\beta$:
$\beta_0\approx0.330458\pi$,
$\beta_*\approx0.330482\pi$.
For $0.25\pi<\beta<\beta_0$, Eqn.\ref{minihx1} describe the minimum contour $k_y^0=\pi/2$ shown in Fig.\ref{finehx1},
for $\beta_*<\beta<0.50\pi$, it describes the C-C$_0$/C-C$_\pi$ boundary.
What happens when $\beta_0<\beta<\beta_*$ is shown in Fig.\ref{finehx2} and summarized below.

 As shown in the Fig.\ref{finehx}, the constant contour line at  $  k^{0}_{y}=\pi/2 $ can be divided into 3 segments:
 (1) $ \pi/4 < \beta < \beta_0 \sim 0.330458\pi $,  $ k^{0}_y =\pm \pi/2 $ is indeed a minimum as shown in Fig.\ref{finehx1},
 (2) $  \beta_0 < \beta < \beta^{*} \sim 0.330482\pi $, $ k^{0}_y =\pm \pi/2 $ becomes a local maximum,
     $  k^{0}_y =0, \pi $ are also local maximum. There are 4 minima $ ( 0, \pm k^{0}_{y} ) $ and
     $ ( 0, \pm (\pi- k^{0}_{y}) ) $ symmetrically located on the two sides of  $ k^{0}_y =\pm \pi/2 $ as shown in Fig.\ref{finehx2}.
     At $ \beta_0 $, the second derivatives of the spectrum at $ k^{0}_y =\pm \pi/2 $  vanish.
  (3) $  \beta^{*} < \beta < \pi/2 $, $ C-C_0 $ and $ C-C_{\pi} $ become two degenerate minima,
      with $ k^{0}_y =\pm \pi/2 $ being still the maximum as shown in Fig.\ref{finehx}.
      At $  \beta^{*} $, the second derivatives of the spectrum at $ k^{0}_y =0, \pi $ vanish.

  So all the two mirror related minima $ ( 0, \pm k^{0}_{y} ) $ and  $ ( 0, \pm (\pi- k^{0}_{y}) ) $
  must end in  the regime  $  \beta_0 < \beta < \beta^{*} $ shown in Fig.\ref{finehx}.

\subsection{ The transition from the  YX-x canted state to the X-FM
at $T=0$ and finite $T$. }

{\sl 1.  The Zero temperature transitions:}

\begin{figure}[!htb]
    \includegraphics[width=0.9\linewidth]{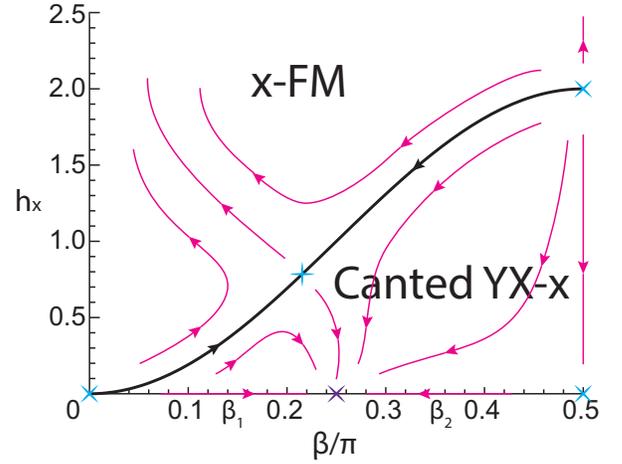}
    \caption{  ( Color online ) The RG flow for RFHM in the transverse field $h_x$. There is a fixed point on the phase boundary controlling
    the zero temperature transition from the YX-x state to the X-FM state which is in the 3d Ising universality class.  At the Abelian point $ \beta=\pi/2 $, the transition is in the 3d XY class. Any deviation from the two  abelian points
    are relevant and drive the system into the 3d Ising fixed point. All the crosses stand for RG fixed
    point.  The whole YX-x state is controlled by the fixed point at $ (\beta=\pi/4, h=0 ) $ which enjoys the enlarged mirror symmetry \cite{rh,rhh}.
    Inside the YX-x state, all the quantum fluctuation generated C-$C_0$, C-$C_{\pi} $ and C-IC are irrelevant under the RG sense, but they
    are the competing to become the driving seed to the X-FM. The Fig.\ref{phasehx},\ref{finehx} show that it is the  C-$C_0$ which wins the competition and becomes the driving seed.}
\label{rghx}
\end{figure}

The transition from the $ YX-x $ canted state to the X-FM is
characterized by the order parameter is $ M_y(T=0)= \langle S^{y} \rangle $.
As said at the beginning of Sec.II, the Hamiltonian Eqn.\ref{hhx} has $ {\cal P}_x $ symmetry:
$ S^{x} \rightarrow S^{x}, k_y \rightarrow - k_y, S^{y} \rightarrow - S^{y},  S^{z} \rightarrow - S^{z} $
and the translational symmetry. The X-FM respects both symmetry, so $ M_y(T=0)=0 $,
but the YX-x states breaks both, but keeps its combination $ {\cal P}_x \times ( x \rightarrow x +1 ) $ as shown in the appendix A,
so $ M_y(T=0) \neq 0 $. Due to the spin-orbital locking, destroying the $ M_y(T=0)= \langle S^{y} \rangle $ order will also
restore the translational symmetry along $ x $ direction.
As shown in  Eqn.\ref{relmag1} and \ref{relmag2}, there are relativistic gapped $ C-C_{0} $ magnons on both  sides indicating the dynamic exponent $ z=1 $.
So we conclude that the transition is in the 3d Ising universality class.
The RG flow is controlled by a fixed point on the phase boundary shown in Fig.\ref{rghx},

At $ h_x=0 $, the $ Y-x $ collinear state is the exact eigenstate \cite{rh}, so $ M_y(T=0) =S $.
The ground state itself contains no information on the $ C-C_0, C-C_{\pi} $ and $ C-IC $ magnons.
As shown in Sec.II-B, any  $ h_x \neq 0 $ transfers the Y-x state into the  $ YX-x $ canted state and also
introduces quantum fluctuations. The canted angle of the classical YX-x state is given in Eqn.\ref{canthx}.
The ground state itself contains information on the $ C-C_0, C-C_{\pi} $ and $ C-IC $ magnons.
They all compete and move to the phase transition boundary.

From the classical YX-x state Eqn.\ref{yxx} with the canted angle Eqn.\ref{canthx} and Eqn.\ref{Hyyx1s}, we find that they reduce
the order parameter below its classical value:
\begin{equation}
  M_y(T=0) = M_c[1 - \frac{1}{2N}\sum_{k} ( \frac{1}{\omega^{+}_{k}} + \frac{1}{\omega^{-}_{k}} )]
\label{yt0}
\end{equation}
  where $ M_c= S \sqrt{ 1- ( h/h_{cx})^{2} } $ is the classical order parameter.
%  So one can see the order parameter reduction comes from both classical reduction and quantum fluctuations.

When approaching the phase boundary $ h_{cx}=2 \sin^2 \beta $, the quantum fluctuations get stronger and stronger,
finally, the $ C-C_0 $ wins the competition, the order parameter vanishes as $ M_y(T=0) \sim ( h_{cx}-h )^{\beta_{3d}} $
with the 3d Ising exponent $ \beta_{3d} \sim 0.31 $.
Eqn.\ref{yt0} leads to $  M_y(T=0) \sim  \Delta \sim (h_{cx}-h)^{1/2} $
with $ \beta_{MF} =1/2 $.  Near the critical line $ h_{cx} $, the $ C-C_0 $ magnon gap $ \Delta $ on both sides own
the critical scaling $  \Delta \sim |h-h_{cx} |^{1/2} $ which gives the mean field exponent$ \nu_{MF} =1/2 $.
Note that $ \nu=0.64 $ for the 3d Ising model. In principle, one may achieve these exact exponents from the
SWE to infinite orders in $1/S $. In practice, one need to find an alternative way such as effective theory
and Renormalization group (RG). The $ T=0 $ RG flow is shown in Fig.\ref{rghx}.

At the Abelian  $ \beta=\pi/2 $ point, staring from $ h > h_{cx} $, as shown in \cite{rhh},
due to the enlarged $ U(1) $ symmetry, the transition is driven by the simultaneous condensations of the two
degenerate minima at $ (\pi,0) $ and $ (\pi, \pi) $, from below $ h < h_{cx} $, it is also a
simultaneous condensations of $ C-C_0 $ and $ C-C_{\pi} $ magnons, so the transition is in the  3d $ XY $  universality class.

{\sl 2. The finite temperature behaviors and transitions }

Except at the abelian point $ \beta=0, h=0 $,
there is a gap $\Delta_{-}(\beta)$ in the excitation spectrum in the YX-x canted phase,
so the order survives up to a finite critical temperature
$ T_{cx} \sim  \Delta_{-}(\beta,h_x) $ above which the system gets to the X-FM state.
At the $ T=0 $ phase boundary in Fig.\ref{phasehx}, $ T_c =0 $.
Note that the spin wave expansion works in the whole phase diagram in Fig.\ref{phasehx} at $ T=0 $, but its use at a finite $ T $ is only limited to
$ T \ll T_{cx} $, it fails when getting too close to $ T_{cx} $.
At low temperatures $ T < T_{cx} $, inside the C-IC regime in Fig.\ref{phasehx}, by expanding  $ \omega_{-}( k ) $ in Eqn.\ref{spectrumhhx}
around the C-IC minima $ (0, \pm k^{0}_y ) $, we find the excitation spectrum takes the relativistic form
\begin{align}
    \omega_q=\sqrt{\Delta_{ic}^2+v_x^2 q_x^2+v_y^2 q_y^2}
    ,\quad
    k=(0, \pm k^{0}_y ) + q
\label{relmagic}
\end{align}
   whose detailed behaviors along the $ k^{0}_y= \pm \pi/2 $ are shown in Fig.\ref{finehx1} and Fig.\ref{finehx2}.
   They dominate the contributions to the magnetization and the specific heat when $ T \ll \Delta_{ic} $:
\begin{eqnarray}
 C_m(T) & \sim & \frac{\Delta^3_{ic}}{2\pi v_xv_y T}e^{-\Delta_{ic}/T}   \nonumber   \\
 M_y(T)  & \sim  &  M_y(T=0)-\frac{T^2}{2\pi v_xv_y}e^{-\Delta_{ic} /T}
\end{eqnarray}
   where $ M_y(T=0) $ is the zero temperature staggered magnetization given in Eq.\ref{yt0}.

Following the procedures in \cite{rh}, one can also evaluate the uniform and staggered susceptibilities along the $ y $ direction,
and various dynamic spin correlation functions. Especially, we expect that the C-IC magnons will lead to
two split peaks located at $ (0, \pm k^{0}_y ) $ in the transverse spin structure factors $ S^{+-}(\vec{k}) $.
All these physical quantities can be measured by specific heat  \cite{heat1,heat2}, In situ measurement \cite{dosexp} and
light or atom Bragg spectroscopy \cite{lightatom1,lightatom2} respectively.

Because  inside the YX-x phase in Fig.\ref{phasehx}, the RG flows to the fixed point $ ( \beta=\pi/4, h=0) $, so
the finite temperature transition from the $ YX-x $ canted phase to the X-FM is in the same universality class as that
at zero field case discussed in \cite{rh}. Its nature remains to be determined in a separate publication \cite{un}.
Of course, at the  Abelian  $ \beta=\pi/2 $ point, it is in the 2d XY universality class.

\section{ Transverse field $ h_z $ }

The RFHM in a transverse field along $S_z$ direction is described by
\begin{align}
    \mathcal{H}
    \!=\!-\!J\!\!\sum_i
    [\mathbf{S}_i R_x(\!\pi\!) \mathbf{S}_{i+x}
    \!\!+\!\mathbf{S}_i R_y(2\beta) \mathbf{S}_{i+y}]
    \!-\!H_z\!\!\sum_i \!S_i^z
\label{hhz}
\end{align}

By applying the hidden $ U(1) $ symmetry operator \cite{rh} at the zero field case, we show that the $ h_z $ case
     can be mapped to the RFHM in a staggered  $ h_x $ field along the $ x $ direction.
     However, as expected the staggered $ h_x $ could make dramatic  difference than the uniform case discussed in the last section.

     Similar to the analysis below Eqn.\ref{hhx}, one can see the Hamiltonian Eqn.\ref{hhz} has the translational symmetry and the
    $ {\cal P}_z $ symmetry: $ k_x \rightarrow - k_x, S^{x} \rightarrow - S^{x}, k_y \rightarrow - k_y, S^{y} \rightarrow - S^{y},
    S^{z} \rightarrow S^{z} $ which is also equivalent to a joint $ \pi $ rotation of both the spin and the orbital  around the $ \hat{z} $ axis.
    It also keeps  $  {\cal T } {\cal P}_x $ and  $  {\cal T } {\cal P}_y $ symmetry.

\subsection{ Z-FM state and excitations in the strong field }

\begin{figure}[!htb]
\includegraphics[width=0.9\linewidth]{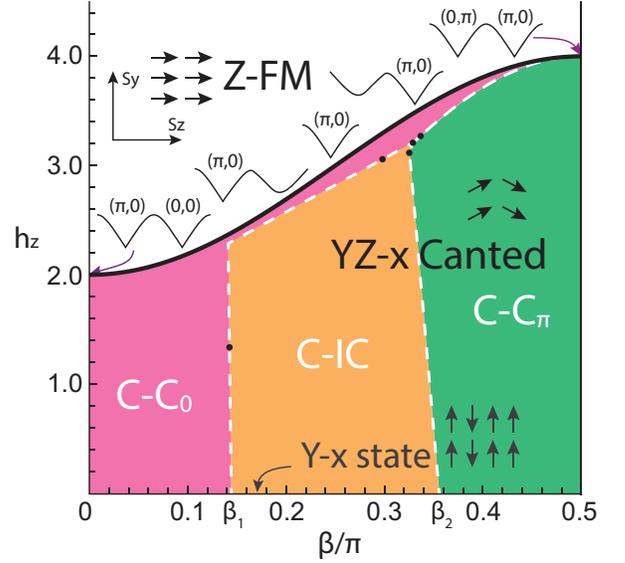}
\caption{ ( Color online ) Phase diagram for RFHM with transverse field $h_z$.
	Boundaries between C-C$_0$,C-C$_\pi$,C-IC are indicated by white dashed lines.
At $h_z=0$, the ground state is the Y-x state which is a exact without any quantum fluctuations.
         Any $ h_z > 0$ will transfer it into the YZ-x  canted state which suffers quantum fluctuations.
         At the critical field $ h_{cz}(\beta) $, the $ YZ-x $ canted state undergoes  a quantum phase transition to the Z-FM state.
         It is  in the 3d Ising university class and always driven by the condensation of $ C-C_0 $ magnons.
         The $ C-IC $ magnons always lose to  $ C-C_0 $ magnons before hitting the phase boundary.
         At the two Abelian points $ \beta=0, \pi/2 $, it is in 3d XY class.
         At the Abelian point $ \beta=\pi/2 $, $ C-C_0 $ and  C-C$_\pi$ magnons condense simultaneously and lead to a
         3d XY class transition to the X-FM.
Due to the lack of generalized Mirror symmetry as in the $h_x $ case, the C-IC regime has a more complicated landscapes than
that in Fig.\ref{phasehx}. The detailed competition in the C-IC regime are shown in Fig.\ref{h3pfine} and Fig.\ref{fineh0b0}.
The 5 dots are explained in appendix C.	}
\label{phasehz}
\end{figure}

In a strong transverse field $H_z\gg J$, the system is in Z-FM phase
with spin classically fully polarized to $S_z$ direction with quantum fluctuations shown in Fig.\ref{phasehz}.
Introducing the HP bosons
$S^+=\sqrt{2S-a^\dagger a}a$,
$S^-=a^\dagger\sqrt{2S-a^\dagger a}$,
$S^z=S-a^\dagger a$,
we map Hamiltonian Eq.\eqref{hhz} to the order of $ 1/S $:
\begin{align}
    \mathcal{H}=&E_0+2JS
    \sum_k[
    (h_z\!-\!2\sin^2\!\beta\!-\!\cos^2\!\beta\cos k_y)
    a_k^\dagger a_k \nonumber\\
    &+(\sin^2\!\beta\cos k_y\!-\!\cos k_x)]
     (a_ka_{-k}+a_k^\dagger a_{-k}^\dagger)/2]
\end{align}
where the classical ground state energy $E_0=2JNS^2\sin^2\beta-H_zNS$
and the dimensionless field $h_z=H_z/(2JS)$.
Now the Hamiltonian can be diagonalized
by a Bogoliubov transformation
\begin{align}
    \mathcal{H}=&E'_0+4JS\sum_k\omega_k\alpha_k^\dagger\alpha_k
\end{align}
where the ground-state energy  at the order of $1/S $ is $E'_0=E_0-2JS\sum_k\omega_k$
and the spin-wave dispersion takes the form£»
\begin{align}
    \omega_k\!\!=\!\!
    \sqrt{\!(h_z\!-\!2\sin^2\!\!\beta\!-\!\cos^2\!\!\beta\cos k_y)^2
	\!\!-\!(\sin^2\!\!\beta\cos k_y\!-\!\cos k_x)^2}
\end{align}
 where, for $0 < \beta < \pi/2$, one can identify there is a unique minimum located at
 $k_0=(k_x,k_y)=(\pi,0)$ with the gap:
\begin{align}
	\Delta_{\pi}=\omega_{k=k^0}=\sqrt{h_z(h_z-3+\cos2\beta)}
\end{align}
and the critical field strength is given by the gap vanishing condition:
\begin{align}
	h_{cz}(\beta)=3-\cos2\beta=2+2\sin^2\beta
\label{hcz}
\end{align}
   which is shown in Fig.\ref{phasehz}

   The Excitation around the minimum takes the relativistic form
\begin{align}
    \omega_q=\sqrt{\Delta^2_{\pi}+v_x^2q_x^2+v_y^2q_y^2},\quad k=k^0+q
\end{align}
where
\begin{align}
    v_x^2&=1+\sin^2\beta,   \nonumber \\
    v_y^2&=(h-1)\cos^2\beta+2\sin^4\beta,
\end{align}
and the critical velocities are $v_{x,c}^2=v_{y,c}^2=1+\sin^2\beta$.
In contrast to the $ h_x $ case, here, the $ v_{x,c} $ and $v_{y,c}$ do not vanish even at $ \beta=0 $.
 The gap vanishing at $k^0=(k_x,k_y)=(\pi,0)$ indicate a quantum phase transition into a spin-orbital correlated state with orbital order
  $ (\pi,0)$. It was known that at $ h_z=0 $, the ground state $ Y-x $ state also has the $ ( \pi,0) $ orbital order.
  That indicates that there is only one phase transition and the state below $ h_{cz} $ could be just the $ YZ-x $ state with a canted angle.
  As to be shown in the next subsection, we show that it is indeed the $ YZ-x $ state with the orbital order $ (\pi,0) $.
  Near the QPT, $ \Delta_{\pi} \sim (h_z - h_{cz} )^{1/2} $.

At the two Abelian points $\beta=0,\pi/2$ the system has SU(2) symmetry in the rotated basis $ \tilde{SU}(2) $
with $\vec{\tilde{S}}_{i} = R(\hat{x},\pi n_1) \vec{S}_{i}$
and $ \tilde{\tilde{SU}}(2) $ with $ \vec{\tilde{S}}_{i} = R(\hat{x},\pi n_1)  R(\hat{y},\pi n_2) \vec{S}_{i} $ respectively.
So Eqn.\ref{hhz} can be mapped to a FM Heisenberg model in $ -h_z \sum_i (-1)^{i_x} \tilde{S}^x_{i} $ ( see Eqn.\ref{tsu2z} ) and
$ -h_x \sum_i (-1)^{i_x +i_y} \tilde{\tilde{S}}^x_{i} $
( see Eqn.\ref{ttsu2z} ) respectively.
So at $ \beta=0, \pi/2 $, taking the result from $ \beta=\pi/2 $, any $ h_z $ will lead to a spin-flop transition resulting into a $ U(1) $ symmetry breaking
canted phase with one Goldstone mode $ \phi $.
Then there is another transition to the Z-FM at a finite $ h_c=2, 4 $ respectively.
These results at the two Abelian points are consistent with the general result Eqn.\ref{hcz}
and shown in Fig.\ref{phasehz}

For $ ( \beta=0, h_z=0 ) $, transferring back from the $ \tilde{SU}(2) $ basis to the original basis,
the Hamiltonian Eq.\eqref{hhz} has the SU(2) symmetry
which is generated by
$\sum_i S_i^x$, $\sum_i (-1)^{i_x}S_i^y$,
and $\sum_i (-1)^{i_x}S_i^z$.
When $h_z>0$, only $\sum_i (-1)^{i_x}S_i^z$ remains as a conserved quantity.
Obviously, the Z-FM state keeps all symmetry from the Hamiltonian.
By acting the conserved quantity $ e^{i \phi \sum_i (-1)^{i_x}S_i^z } $ on the minima $(\pi,0)$ will generate
another minimum at $ (0,0) $. So we conclude that at  $ ( \beta=0£¬ h_z>0 ) $,
 the system has two minima located at $(0,0)$ and $(\pi,0)$ as shown in Fig.\ref{phasehz}.

For $ (\beta=\pi/2, h_z=0 ) $,
transferring back from the $ \tilde{\tilde{SU}}(2) $ basis to the original basis, the Hamiltonian Eq.\eqref{hhz} has the SU(2) symmetry
which is generated by $\sum_i(-1)^{i_y} S_i^x$, $\sum_i (-1)^{i_x}S_i^y$,
and $\sum_i (-1)^{i_x+i_y}S_i^z$.
When $h_z>0$, only $\sum_i (-1)^{i_x+i_y}S_i^z$ remain as a conserved quantity.
Acting the conserved quantity $ e^{i \phi \sum_i (-1)^{i_x+i_y}S_i^z } $ on the minima $(\pi,0)$,
generates another minimum at  $(0,\pi)$. So we conclude that at $ ( \beta=\pi, h_z>0 ) $,
the system has two minima located at $(\pi,0)$ and $(0,\pi)$  as shown in Fig.\ref{phasehz}.

\subsection{ The Co-planar YZ-x Canted state below $h_{cz}$: }

{\sl 1. Classical YZ-x canted phase at $h<h_{cz}$ }

(a) Approaching from the left Abelian point $ \beta=0 $.

At $ \beta=0 $, in the  $ \tilde{SU}(2) $ basis $\tilde{\tilde{\mathbf{S}}}_i=R_x(i_x\pi)\mathbf{S}_i$,
the Hamiltonian Eq.\ref{hhz} takes the form:
\begin{align}
	\mathcal{H}=-J\sum_{\langle ij\rangle}
	\tilde{\mathbf{S}}_i\cdot\tilde{\mathbf{S}}_j
	-H_z\sum_i(-1)^{i_x}\tilde{S}_i^z
\label{tsu2z}
\end{align}
   When $0<H_z<H_{zc}$ the classical state in the $ \tilde{SU}(2) $ basis is:
\begin{align}
    \tilde{\mathbf{S}}_i=
    S\big(\sin\theta\cos\phi,\sin\theta\sin\phi,(-1)^{i_x}\cos\theta\big)
\label{statez10}
\end{align}

  Reverting back to original basis leads to the classical state in original basis:
\begin{align}
    \mathbf{S}_i=
    S\big(\sin\theta\cos\phi,(-1)^{i_x}\sin\theta\sin\phi,\cos\theta\big)
\label{statez20}
\end{align}

  Although we obtained Eq.\eqref{statez10} and Eq.\eqref{statez20}
at $\beta=0 $, the same ansatz hold for $ 0<\beta<\pi/2 $ whose classical ground energy is:
\begin{align}
	E_{\rm c}
	=-2NJS^2[1-(1+\sin^2\beta)\cos^2\theta+h_z\cos\theta\nonumber\\
	-\sin^2\beta\sin^2\theta\cos^2\phi]
\label{classez}
\end{align}
 It is easy to see that any $ \beta >0  $ explicitly breaks the $ U(1) $ symmetry at $ \beta=0 $,
 so picks up $\phi=\pi/2 $ and leads to the classical YZ-x canted state:
\begin{align}
    \mathbf{S}_i=S\big(0,(-1)^{i_x}\sin\theta,\cos\theta\big)
\label{yzx}
\end{align}
  with the corresponding classical ground state energy
\begin{align}
	E_{\rm c}=&
	-2NJS^2[1-(1+\sin^2\beta)\cos^2\theta+h_z\cos\theta]
\label{yzxen}
\end{align}

  Minimization of Eq.\eqref{yzxen} leads to the canted angle:
\begin{align}
    \cos\theta=\frac{h_z}{2(1+\sin^2\beta)}<1,
    \quad \textmd{when } h_z<h_{cz}
\label{canthz}
\end{align}
   which always has a solution as long as $ h_z<h_{cz} $.

(b) Approaching from the right Abelian point $ \beta=\pi/2 $.

 In fact, one can reach the same results in Eqn.\ref{yzx} and \ref{yzxen} from the right at $ \beta=\pi/2 $.
 In the $ \tilde{\tilde{SU}}(2) $ basis $\tilde{\mathbf{S}}_i=R_x(i_x\pi)R_y(i_y\pi)\mathbf{S}_i$,
 the Hamiltonian  in Eq.\eqref{hhz} at $ \beta=\pi/2 $ takes the form:
\begin{align}
    \mathcal{H}=-J\sum_{\langle ij\rangle}
	\tilde{\tilde{\mathbf{S}}}_i\cdot\tilde{\tilde{\mathbf{S}}}_j
	-H_z\sum_i(-1)^{i_x+i_y}\tilde{\tilde{S}}_i^z
\label{ttsu2z}
\end{align}

When $0<H_z<H_{cz}$, the classical ground state is:
\begin{align}
    \tilde{\tilde{\mathbf{S}}}_i=
    S\big(\sin\theta\cos\phi,\sin\theta\sin\phi,(-1)^{i_x+i_y}\cos\theta\big)
\end{align}

Reverting back to the original basis leading to the  classical ground state in the original basis
\begin{align}
    \mathbf{S}_i=
    S\big((-1)^{i_y}\sin\theta\cos\phi,
      (-1)^{i_x}\sin\theta\sin\phi,
      \cos\theta\big)
\label{yzxR}
\end{align}
  with the classical ground state energy
\begin{align}
	E_{\rm c}
	=-2NJS^2[1-(1+\sin^2\beta)\cos^2\theta+h_z\cos\theta \nonumber\\
	 -\cos^2\beta\sin^2\theta\cos^2\phi]
\label{yzxenR}
\end{align}
  Obviously, any $\beta<\pi/2$  picks up $\phi=\pi/2$. Then Eqn.\ref{yzxR} and Eqn.\ref{yzxenR} reduce to
  Eqn.\ref{yzx} and Eqn.\ref{yzxen} respectively.

{\sl 2. Spin wave analysis in the YZ-x Canted state }

  Starting from the classical YZ-x state Eqn.\ref{yzx} and using similar procedures to obtain Eqn.\ref{YXxen1s}
  we obtain the spin-wave dispersion:
\begin{widetext}
\begin{align}
    \omega_k^{\pm}\!\!=\!\!
    \sqrt{A_k^2\!+\!B_k^2\!+\!C^{\prime 2}_k\!-\!C^{\prime \prime 2}_k\!-\!D_k^2
    \!\pm\!
    2\sqrt{(A_k^2\!-\!D_k^2)B_k^2\!+\!(A_k C^{\prime}_k\!-\!C^{\prime \prime}_k D_k)^2}}
\label{YZxen1s}
\end{align}
\end{widetext}
where the expressions of $ A_k, B_k, D_k $ are listed in Eqn.\ref{abcdhx} and
\begin{align}
    C^{\prime}_k&=\sin^2\theta\cos k_x     \nonumber \\
    C^{\prime \prime}_k&=\cos^2\theta\cos k_x 
\end{align}
  where one can see $ C^{\prime}_k + C^{\prime \prime}_k = C_k= \cos k_x $  listed in Eqn.\ref{abcdhx}.
  Of course, the $\theta$ in Eqn.\ref{canthz} in the $H_z$ field is different from that in Eqn.\ref{canthx} in the $H_x$ field.

 From Eqn.\ref{YZxen1s}, one can determine the minimum positions inside the YZ-x state.
 The general structure of Fig.\ref{phasehz} is similar to the $ h_x $ case Fig.\ref{phasehx}.
 However, due to the lack of generalized mirror symmetry as in the $ h_x $ case,
 the detailed landscape of the C-IC regime in Fig.\ref{phasehz} is much more complicated than that in the $ h_x $ case.
 In this subsection, we only outline the general structure. In the next subsection and appendix C,
 we describe details of the shape of the C-IC regime in Fig.\ref{phasehz}.

In Fig.\ref{phasehz}, we still found there are three regimes inside the YZ-x Canted state:
C-C$_{0}$ regime, C-IC regime, and  C-C$_{\pi}$ regime which, at $ h_x=0 $, reduce to
the three regimes identified in \cite{rh}.
Among the three magnons, only C-C$_{0}$ wins the game and drives the transition,
so the transition from the YZ-x state to the Z-FM is driven by the condensations of the C-C$_{0}$ magnons only.
The C-IC magnons still loses to the C-C$_{0}$ in the competition.

Now we can check the consistence of the orbital orders on both sides of $ h_{cz} $.
The YZ-x state has the orbital order $ (\pi,0) $, the $ C-C_0 $ has the orbital order $ (0,0)=(\pi,0) $ in the RBZ.
So its condensation on the top of  YZ-x could lead to the two orbital orders either $  (\pi,0)+(0,0)=(\pi,0) $
or $  (\pi,0)+(\pi,0)=(0,0) $ in the EBZ. The $ (0,0) $ order  is nothing but that of the Z-FM in Fig.\ref{phasehz}

\subsection{ Fine structure of the $ C-IC $ magnons inside the $ C-IC $ regime in Fig.\ref{phasehz}. }

\begin{figure}[!htb]
\includegraphics[width=0.95\linewidth]{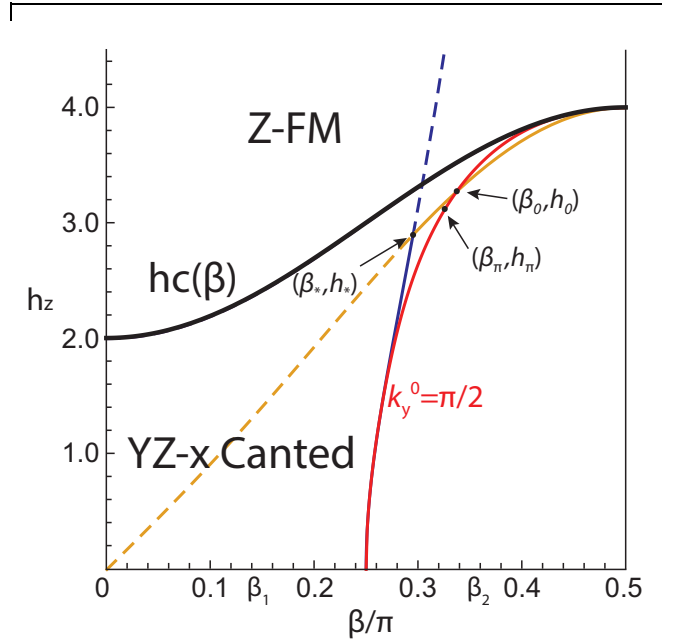}
\caption{  ( Color online ) 
The three backbone lines $h_2$ ( brown ) , $h_3$ ( blue ), $h'_3$ ( red ) and their crossings in the $ h_z $ case.
Thick solid line is the phase boundary $ h_{cz} $.
The  $h'_3$ is the constant contour at $ k^{0}_{y}=\pm \pi/2 $.
Along the solid line parts of $h_2$ and $h_3$, $ C-C_0 $ and $ C-C_{\pi} $ become degenerate $\omega(k_x=0,k_y=0)=\omega(k_x=0,k_y=\pi)$.
The dashed parts of $h_2$ and $h_3$ are extraneous solutions.
 Due to the lack of generalized mirror symmetry as in the $ h_x $ case in Fig.\ref{finehx}, one need the 3 separate lines $ h^{\prime}_3 $ and $h_2, h_3 $
 to describe the competitions of the three kinds of magnons.
 The $ h^{\prime}_3 $ can be split into two segments $ \pi/4 < \beta < \beta_{\pi} $, $ (0, k^{y}_{0}=\pm \pi/2 ) $
  is the minimum position. $ \beta_{\pi}  < \beta <  \beta_{0} $, $ C-C_{\pi} $ becomes the minimum position.
  When  $ \beta_{0}  < \beta < \pi/2 $, $ h^{\prime}_3 $ rises above $ h_2 $, moves into the $ C-C_{0} $ regime. }
\label{h3pfine}
\end{figure}

  As shown in appendix C, the line $ h^{\prime}_3 $ in Fig.\ref{h3pfine} is determined by setting the first derivative
  of dispersion vanishing at $ (0, k^{y}_{0}=\pi/2 ) $.
  The line $ h_2 $ and $ h_3 $ are determined by the condition that $ C-C_{0} $ and $ C-C_{\pi} $ become degenerate.
  There is one crossing point $ ( \beta_0, h_0) $ between $ h^{\prime}_3 $ and $ h_2 $  in Fig.\ref{h3pfine}.

  In the $ h_x $ case discussed in Sec.II, both conditions are the same, so lead to just one single line
  with the 3 different segments in Fig.\ref{finehx} presented in Sec.II-C.
  However, in  the $ h_z $ case, there are two different conditions which leads to three different lines $ h^{\prime}_3 $ and $ h_2, h_3 $,
  which make the detailed shape of the C-IC regime more complicated than that in $ h_x $ case.

  Along the $ h^{\prime}_3 $, the minimum at $ k^{y}_{0}=\pi/2 $, stays as the  ( local ) minima until
  $ \beta_{flat} \sim 0.33 \pi $ where the second derivative of the
  dispersion  at $ (0, k^{y}_{0}=\pi/2 ) $ vanishes, then it becomes a maximum after $ \beta > \beta_{flat} $.
  ( In fact, before getting to $ \beta_{flat} \sim 0.33 \pi $, there is another point ( let's call it $ \beta_{\pi} $ in Fig.\ref{h3pfine} )
  where the $ (0, k^{y}_{0}=\pi/2 ) $ is just a local minimum, while the $ C-C_{\pi} $ becomes the global minimum ).
  Then $ C-C_{\pi} $ becomes the minimum, while $ C-C_{0} $ becomes the maximum, then until $ C-C_{\pi} $ and $ C-C_{0} $ becomes degenerate
  at $ \beta_0 \sim 0.33729 \pi $ in Fig.\ref{h3pfine}.
   After $ \beta > \beta_0 $, it moves into the $ C-C_0 $ regime where $ C-C_0 $ becomes the minimum.
  $ h^{\prime}_3 $ rises above $ h_2 $ line. So $  \beta_0 \sim 0.33729 \pi $ is determined by setting $ h_2=h^{\prime}_3 $ as shown in Fig.\ref{h3pfine}.

  So in practice, the $ h^{\prime}_3 $ can be split into two segments $ \pi/4 < \beta < \beta_{\pi} $, $ (0, k^{y}_{0}=\pi/2 ) $
  is the minimum position. $ \beta_{\pi}  < \beta <  \beta_{0} $, $ C-C_{\pi} $ becomes the minimum position.
  ( So $ \beta_{flat} $ is really not that important anymore ).
  Then when  $ \beta_{0}  < \beta < \pi/2 $, $ h^{\prime}_3 $ rises above $ h_2 $, moves into the $ C-C_{0} $ regime.
  Then we have to use the $ h_2 $ line to delineate the  $ C-C_0 $ and $ C-C_{\pi} $ boundary.

  So the $ C-IC $ boundary along $ (0, k^{y}_{0}=\pi/2 ) $ happens at $ ( \beta_{\pi}, h_{\pi} ) $ where it turns into $ C-C_{\pi} $.
 In principle, one can determine the whole  $ C-IC $ boundary in the whole $ YZ-x $ phase.
  Indeed, we determine the $ C-IC $ boundary along the line $ h_2 $ and $ h_3 $ in Fig.\ref{h2h3}.
  Connecting all the special points along the three lines $h_2, h_3, h'_3 $ in Fig.\ref{h3pfine} and Fig.\ref{h2h3} in the appendix C.
  and also $ \beta_1, \beta_2 $ at $ h_x=0 $ lead to Fig.\ref{phasehz}
  and also the evolution around $(\beta_0,h_0) $ in Fig.\ref{fineh0b0}.

%  Unfortunately, due to these complicated landscapes due to the multiple minima distributions, it is quite difficult to determine the complete
%  regimes of $ C-C_0, C-C_{\pi} $ and $ C-IC $ very precisely.
%  So we refrain from drawing them in the figure.
%   We are assured that the $ YZ-x $ state always gets to
%   the $ Y-x $ state through the $ C-C_0 $, so there are always a $ C-C_0 $ regime below the transition line.

\subsection{ The transition from the $ YZ-x $ canted state to the Z-FM at $ T=0 $ and finite $ T $. }

{\sl 1. The $ T=0 $  transitions }

The transition from the $ YZ-x $ canted state to the Z-FM at $ T=0 $ is still
characterized by the order parameter  $ M_y(T=0)= \langle S^{y} \rangle $.
As said at the beginning of Sec.III, the Hamiltonian Eqn.\ref{hhx} has
has the translational symmetry and the
    $ {\cal P}_z $ symmetry: $ k_x \rightarrow - k_x, S^{x} \rightarrow - S^{x}, k_y \rightarrow - k_y, S^{y} \rightarrow - S^{y},
    S^{z} \rightarrow S^{z} $.
 The Z-FM respects both symmetry, so $ M_y(T=0)=0 $,
but the YZ-x states breaks both, but still keeps the combination  $ {\cal P}_z \times ( x \rightarrow x+1 ) $, so $ M_y(T=0) \neq 0 $.
Due to the spin-orbital locking, destroying the $ M_y(T=0)= \langle S^{y} \rangle $ order will also
restore the translational symmetry along $ x $ direction.
Similar to the $ h_x $ case, there are relativistic gapped $ C-C_{0} $ magnons on both  sides indicating the dynamic exponent $ z=1 $.
So we conclude that the transition is also in the 3d Ising universality class.
%The RG flow is controlled by a fixed point on the phase boundary shown in Fig.\ref{rghx}.
The LSWE only leads to the mean field exponent $ \beta_{MF}=1/2, \nu_{MF}=1/2 $.

At the two  Abelian  points $ \beta=0 $ ( or $ \beta=\pi/2 $ ), starting from $ h > h_{cz} $, as shown in \cite{rhh},
due to the enlarged $ U(1) $ symmetry, the transition is driven by the simultaneous condensations of the two
degenerate minima at $ (0,0) $ and $ (\pi, 0) $ ( or $ (0,\pi) $ and $ (\pi,0) $ ) shown in Fig.\ref{phasehz} and  is in the universality class of 3d $ XY $ model.
From below $ h < h_{cx} $, at $ \beta=0 $, it is just the condensation of $ C-C_0 $ magnons,
at $ \beta=\pi/2 $, it is a simultaneous condensations of $ C-C_0 $ and $ C-C_{\pi} $ magnons,
so the transition is also in the 3d $ XY $  universality class.
After considering the above differences, the $ T=0 $ RG flow diagram is similar to Fig.\ref{rghx}.

{\sl 2. The finite temperature behaviors and transitions }

Because  inside the YZ-x phase in Fig.\ref{phasehz}, the RG flows to the fixed point $ ( \beta=\pi/4, h_z=0) $, so
the finite temperature transition at $ T_{cz} $ from the $ YZ-x $ canted phase to the Z-FM is in the same universality class as that
at zero field case discussed in \cite{rh}. Its nature remains to be determined.
Of course, at the two  Abelian points  $ \beta=0, \pi/2 $, it is in the 2d XY universality class.
At the $ T=0 $ phase boundary in Fig.\ref{phasehz}, $ T_c =0 $.
All the physical quantities at $ T  \ll T_{cz} $ can be similarly evaluated as in $ h_x $ case.

\section{Discussions and conclusions}

The C-IC magnons in the zero field RFHM stand for short-ranged In-commensurate orders embedded in a long-ranged ordered commensurate phase \cite{rh}.
In order to transfer the short-ranged In-commensurate orders into long-range ordered ones,
one need to drag out these C-IC and then drive them into condensations.
However, as shown in \cite{rhh} and this work,  these C-IC response quite differently to the $ h_y $ and $ h_x, h_z $ field.
In the $ h_y $ case, at a small $ h_y < h_{c1} $, the $ Y-x $ state stay as the exact ground state, so C-$C_0$, C-$C_{\pi} $ and C-IC
remain extrinsic, detached form the exact ground state and need to be thermally excited.
As $ h_y \rightarrow h^{-}_{c1} $, the C-IC always emerge as the driving seeds to lead to various IC-SKX phase through a line of
fixed points at $ h_y = h^{-}_{c1} $.
However, in both $ h_x $ and $ h_z $ case, the C-IC always lose to C-$C_{0} $ which is the driving seeds to lead to X-FM and Z-FM respectively.
In fact, one can also group $ h_y $ and $ h_x $ as an in-plane field, while $ h_z $ as the perpendicular field.
In the in-plane case, there is Mirror symmetry or a generalized mirror symmetry respectively to characterize the competition among the magnons.
While in the perpendicular field, there is no such mirror symmetry.

In the $ (\beta, h_y ) $ phase diagram, the IC-SKX phase is surrounded by 4 other phases:
the two commensurate co-planar canted phases at the left and right in the SOC parameters, two collinear phases in the low and high field.
The two canted phases and the Ic-SKX phases break the $ U(1) $ symmetry spontaneously, so support a gapless excitation.
The transition from the canted phase to the Y-FM in the high field is in the 3d XY transition class, controlled by the
RG fixed point at the two Abelian points. This is in sharp contrast to the YX-x ( YZ-x ) canted phase to 
the high field X-FM ( Z-FM £© in Fig.\ref{phasehx} ( Fig.\ref{phasehz} ) which is in the 3d Ising transition class,
controlled by the RG¡¡fixed point in the middle of the phase boundary ( see Fig.\ref{rghx} for the $ h_x $ case )
instead of at the two Abelian points.
As stressed in \cite{rhh} and appendix A,
in the $ h_y $ case, the IC-SKX is due to the condensations of non-relativistic C-IC at a single minim $ ( 0, k^{0}_y ) $
from $ h < h_{c1} $, so the transition has the dynamic exponent $ z=2 $.
On the experimental side, the IC-SKX phase match rather naturally and precisely
the incommensurate, counter-rotating (in A/B sublattice), non-coplanar magnetic orders detected on iridates
$\alpha,\beta,\gamma$-Li$_2$IrO$_3$ \cite{kitpconf}.
Both $ h_x $ and $ h_z $ explicitly breaks the spin-orbital coupled $ U(1) $ symmetry of the RFHM at a zero field.
The YX-x or YZ-x phase supports only gapped magnons.
Unfortunately, as shown in Fig.\ref{phasehx} and Fig.\ref{phasehz},
the relativistic C-IC with at least two minima at $ ( 0, \pm k^{0}_y ) $ always lose to $ C-C_0 $, so can not emerge to drive any phase transitions.
There is only one  transition which is driven by the condensation of $ C-C_0 $ and is in the 3d Ising universality class.
Of course, the Finite temperature transitions in $ h_y $ and $ h_x,h_z $ cases are also quite different.

 It is easy to see why the transition from YX-x to X-FM in Fig.\ref{phasehx} and YZ-x to Z-FM  in \ref{phasehz} have to go through
 $ C-C_0 $ instead of $ C-C_{\pi} $. This is because YX-x or YZ-x  have the orbital order $ (\pi,0) $,
 the $ C-C_0 $ has the orbital order $ (0,0)=(\pi,0) $ in the RBZ.
 So its condensation on the top of  YX-x or YZ-x order  could lead to two orbital orders either $  (\pi,0)+(0,0)=(\pi,0) $
 or $  (\pi,0)+(\pi,0)=(0,0) $ in the EBZ. The $ (0,0) $ order  is nothing but that of the X-FM in Fig.\ref{phasehx} or Z-FM  in \ref{phasehz}.
 However, the $ C-C_{\pi} $ has the orbital order $ (0,\pi)=(\pi,\pi) $ in the RBZ.
 So its condensation on the top of  YX-x or YZ-x order  could lead to two orbital orders either $  (\pi,0)+(0,\pi)=(\pi,\pi) $
 or $  (\pi,0)+(\pi,\pi)=(0,\pi) $ in the EBZ, none of the two contains the $ (0,0) $ order.

It is instructive to compare the C-IC magnons with quantum fluctuations generated vortices in $ p/q $ filling Boson Hubbard models
\cite{honey1,dual1,dual2,dual3}, those in high $ T_c $ superconductors \cite{hightc1,hightc2} and exciton superfluids in Bilayer or trilayer
quantum Hall systems \cite{blqh,tri}.
The vortices are gapped topological excitations inside a superfluid, there are at least $ q $ degenerate minima in their dispersions
which transform to each other under the projective representation of the Magnetic space group ( MSG).
So the gap closing ( or condensations ) of the $ q $ minima lead to various kinds of lattice symmetry breaking insulating states.
So these quantum fluctuations generated vortices are short-range translational symmetry breaking insulating orders embedded inside
the translational invariant superfluid states. Even inside the superfluid state, they are the crucial ingredients of the superfluid ground state
and are generated by the intrinsic quantum fluctuations. Their condensations tuned by interactions
spark quantum phase transitions into various neighboring insulating states breaking various translational symmetries of lattices.
Of course, vortices are bosons and satisfy boson statistics.
Here, these $ C-C_{0} $ and $ C-IC $ gapped magnons inside the YX-x or YZ-x state play similar roles as the vortices inside a translational
invariant superfluid state.  They are the crucial ingredients of the YX-x or YZ-x state
and are generated by the intrinsic quantum fluctuations. Their condensations tuned by various Zeeman fields
spark quantum phase transitions into various neighboring spin-orbital correlated commensurate or in-commensurate phases.
 The salient feature of the C-IC magnons is that they may condense at any in-commensurate wavevector leading to incommensurate spin-orbital correlated
 magnetic phases.  This indeed what happens in the $ h_y $ Zeeman field studied in \cite{rhh}. However, in the $ h_x, h_z $ fields studied in this paper,
 they are eliminated before their possible condensations.

 In a recent preprint \cite{rafhm}, we studied Rotated Anti-ferromagnetic Heisenberg model ( RAFHM) which is the fermionic analog of the RFHM \cite{rh}.
 We found that the $ C-C_0, C-IC $  magnons in the RAFHM are also intrinsic ones generated by quantum fluctuations,
 take relativistic dispersion and already embedded in the ground state.
 Their parameters such as the minimum positions $ ( 0, \pm k^{0}_y ) $, gap, velocities $ v_x, v_y $ can be precisely
 measured by the peak positions, the width and Lorentizan shape of the transverse structure factor at $ T=0 $ respectively.
 In this sense, the relativistic $ C-C_0, C-IC $  magnons in the Y-y state in the RAFHM at zero field resemble 
 those in the YX-x and YZ-x canted state studied in this paper.

 The multiple local ( meta-stable ) or global minima structure of the C-IC magnons shown in Fig.\ref{finehx1},\ref{finehx2},\ref{h3pm}
 indicate some short-ranged quantum fluctuations with multiple length scales. These complex structure is intrinsic and
 embedded in the quantum ground state, which may resemble the complex multiple local minimum landscapes in quantum spin glass \cite{glass1,glass2,glass3,glass4}. However, the former is SOC induced, the latter is due to quenched disorders. So the SOC may induce some similar complex phenomena as the disorders.
 
 In this work, we only focus along $ (\alpha=\pi/2, \beta ) $. Obviously, it is important to study how these magnons response when $ \alpha \neq \pi/2 $.
 We expect that turning on $ \alpha -\pi/2 $ will lead to new competitions different from all the three Zeeman field cases.

 In this work, we only focus on quantum phases with only bosonic excitations and without topological orders.
 As said in the introduction, in fermionic systems \cite{blqh,tri,tqpt,ssdw,pairing,rafhm},
 the quantum phase supports both fermionic excitation and collective bosonic excitations.
 The two sectors may compete to lead to various other quantum phases under various external probes.
 In a recent preprint \cite{pairing2}, we studied the attractive Hubbard model with Rashba or Dresselhaus spin-orbit coupling
 in a 2d square lattice subject to a perpendicular  $ h_z $ field which is the weak coupling and negative interaction cousin of Eqn.\ref{hhz}.
 We find it is  the touching ( or gap closing ) of fermionic quasi-particle excitations which signify a topological transition
 from a topological SF to a trivial one or to a band insulator. Obviously, a fermionic quasi-particle can not condense, but they could
 change the topological winding numbers, therefore spark topological transitions.
 It remains much more challenging to study topological transitions driven by condensations of fractionized particles satisfying
 Abelian or non-Abelian  statistics \cite{wenbook,frad}. Unfortunately, in contrast to bosonic or fermionic excitations,
 one may not be able to treat these fractionized particles as independent particles due to their long-range
 entanglements meditated by Abelain or non-Abelian Chern-Simon interactions \cite{wenbook,frad,hightc1,qh,qhdis}.

{\bf Acknowledgements}

We acknowledge  NSF-DMR-1161497, NSFC-11174210 for supports. The work at KITP was supported by NSF PHY11-25915.
W.M. Liu is supported by NSFC under Grants No. 10934010 and No. 60978019, the NKBRSFC under Grants No. 2012CB821300.

\appendix

\section{ Energy spectrum symmetry analysis of $ YX-x $ state in $ h_x $ case and $ YZ-x $ state in $ h_z $ case. }

\begin{figure}[!htb]
	%\centering
	\includegraphics[scale=0.55]{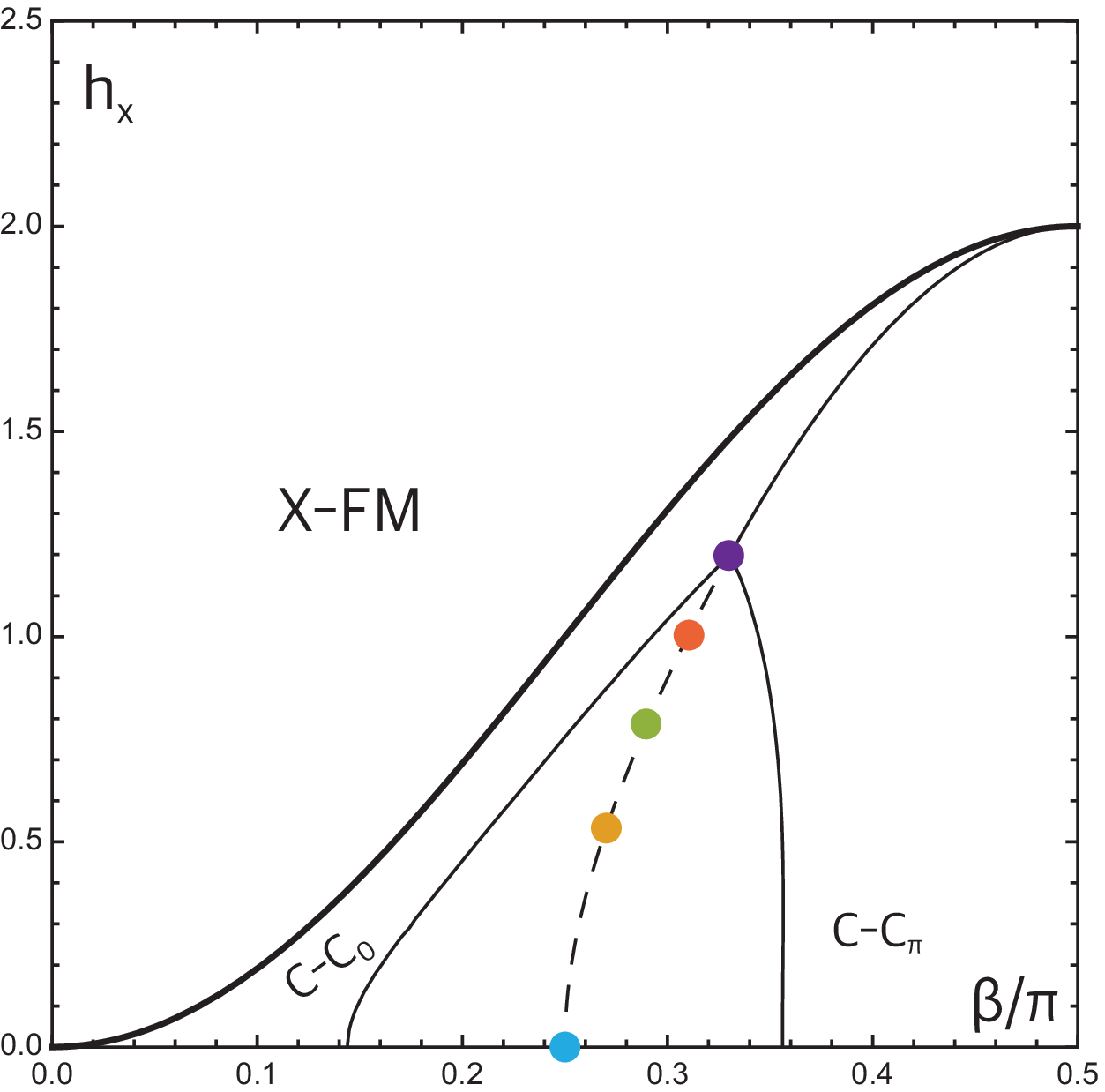}
	\quad
	\includegraphics[scale=0.7]{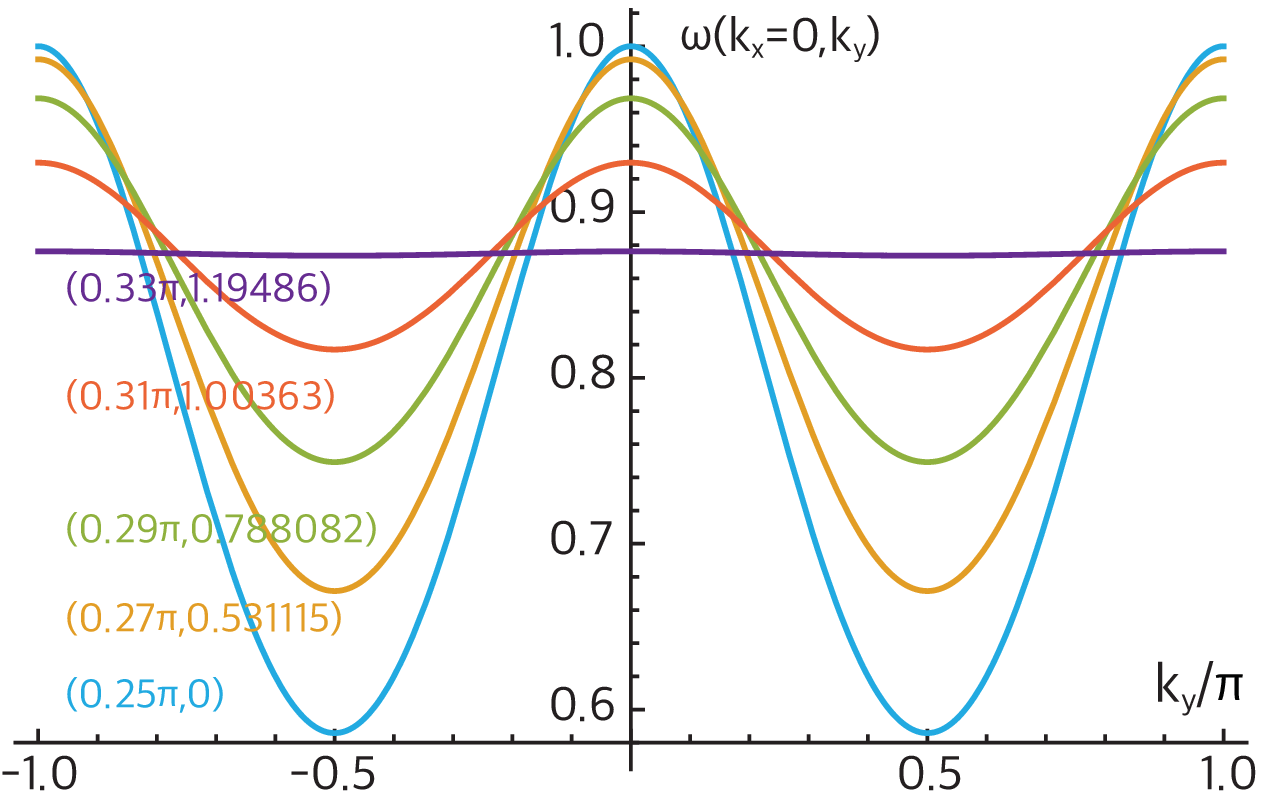}
	\caption{  ( Color online ) (Top) Phase diagram in the $ h_x $ case.
    The dashed line indicates the contour $k_y^0= \pm \pi/2$ inside the C-IC regime.
	Colored dots stand for the 5 parameters  $(\beta,h(\beta))$:
	$(0.25\pi,0)$, $(0.27\pi,0.531115)$, $(0.29\pi,0.788082)$,
	$(0.31\pi,1.00363)$, $(0.33\pi,1.19486)$ used for the figure below falling in the range
    $ \pi/2 < \beta < \beta_0 \approx 0.330458\pi$.
	(Bottom) Spin wave spectrum corresponding to the 4 dots in the top figure.
    There is no other contours crossing the $k_y^0=\pi/2$  contour in this range.
	For example, the contour $k_y^0=\pi/2+10^{-5}$ will hit the extension of the contour $k_y^0=\pi/2$
    in the range $ \beta_0\leq\beta\leq\beta_*$ ( see Fig.\ref{finehx2} ).
    As $ h_x $ increases, the gap at $ k_y^0= \pm \pi/2$ increases.
	Note that even at $ \beta=0.33\pi, h=1.19486 $, the spectrum is still not a straight line. 	}
\label{finehx1}
\end{figure}

\begin{figure}[!htb]
	%\centering
	\includegraphics[scale=0.5]{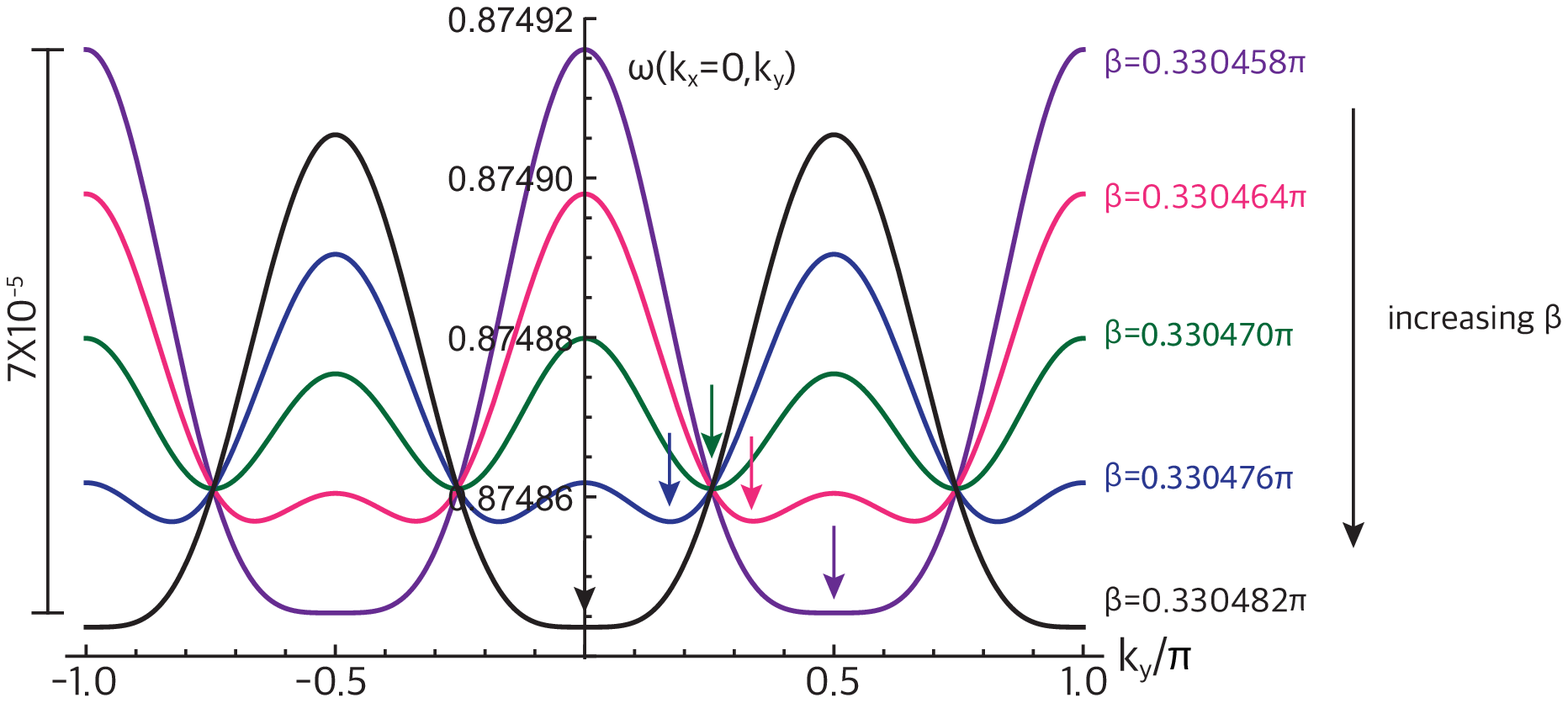}
	\caption{  ( Color online )  The Spin-wave spectrum for several $(\beta,h(\beta))$
	in the range $\beta_0\leq\beta\leq\beta_*$ in the top part of Fig.\ref{finehx1}.
	Arrow is the guide to see the minima locations. Due to the generalized mirror symmetry,
	there are 4 degenerate minima  which are symmetric with respect to $k_y= \pm \pi/2$ respectively.
    Due to the $ k_y \rightarrow - k_y $ symmetry, the 4 minima are also  symmetric with respect to $k_y=0$.
	As increasing $\beta$ from $\beta_0$ to $\beta_*$,
	the minima positions continuously shift from $\pi/2$ to either $0$ or $\pi$.
	For the red curve ($\beta=0.330464\pi$), the 4 minima are at
	$k_y^0=\pm\pi/3,\pm2\pi/3$. It also indicates the 4 contours will
	hit the extension of the contour $k_y^0=\pm \pi/2$ at $\beta=0.330464\pi$ shown in Fig.\ref{finehx}.	}
\label{finehx2}
\end{figure}

As said at the beginning of Sec.II, the RFHM in the $ h_x $ transverse field Eqn.\ref{hhx} enjoys the
$ {\cal P}_x $ symmetry:
$ S^{x} \rightarrow S^{x}, k_y \rightarrow - k_y, S^{y} \rightarrow - S^{y},  S^{z} \rightarrow - S^{z} $ and the translational symmetry.
The$ YX-x $ state breaks both the $ {\cal P}_x $ symmetry and the translational symmetry by one lattice site $ ( x \rightarrow x+1 ) $, but
keeps the combination of the two $ {\cal P}_x \times ( x \rightarrow x+1 ) $. So the excitation spectrum  must have the
$ k_y \rightarrow - k_y $ symmetry. This is indeed respected by the LSW spectrum shown in Fig.\ref{finehx}, \ref{finehx1} and \ref{finehx2}.

Very similarly, as said at the beginning of Sec.III, the RFHM in the $ h_z $ transverse field Eqn.\ref{hhz} enjoys the translational symmetry
and the  $ {\cal P}_z $ symmetry: $ k_x \rightarrow - k_x, S^{x} \rightarrow - S^{x}, k_y \rightarrow - k_y, S^{y} \rightarrow - S^{y},
S^{z} \rightarrow S^{z} $ which is also equivalent to a joint $ \pi $ rotation of both the spin and the orbital  around the $ \hat{z} $ axis.
The$ YZ-x $ state breaks both the $ {\cal P}_z $ symmetry and the translational symmetry by one lattice site $ ( x \rightarrow x+1 ) $, but
keeps the combination of the two $ {\cal P}_z \times ( x \rightarrow x+1 ) $. So the excitation spectrum  must have the
$ k_y \rightarrow - k_y $ symmetry also. This is indeed respected by the LSW spectrum shown in Fig.\ref{h3pm} and \ref{h2h3}.

The zero field RFHM studied in \cite{rh} has the translational symmetry and the  {\cal T}, $ {\cal P}_x, {\cal P}_y $ and  $ {\cal P}_z $ symmetry.
The Y-x state breaks all these symmetries except the $ {\cal P}_y $, however, it still keeps $ {\cal P}_x \times ( x \rightarrow x+1 ) $ and
 $ {\cal P}_z \times ( x \rightarrow x+1 ) $ symmetry, so the excitation spectrum  must have the
$ k_y \rightarrow - k_y $ symmetry also, as indeed respected by the LSW spectrum  shown in \cite{rh}
However, the  energy spectrum in the longitudinal $ h_y $ field studied in \cite{rhh} has no such  $ k_y \rightarrow - k_y $ symmetry anymore.
The RFHM in the longitudinal $ h_y $  field  enjoys the translational symmetry and the
$ {\cal P}_y $ symmetry: $ {\cal P}_y $ symmetry: $ S^{y} \rightarrow S^{y}, k_x \rightarrow - k_x, S^{x} \rightarrow - S^{x},  S^{z} \rightarrow - S^{z} $.
The $ Y-x $ state keeps $ {\cal P}_y $ symmetry, but breaks the translational symmetry by one lattice site $ ( x \rightarrow x+1 ) $,
So the excitation spectrum may not have the $ k_y \rightarrow - k_y $ symmetry.
Indeed, the $ h_y $ field will just pick one the two degenerate minima $ \pm k^{0}_y $ and condense it at $ h=h_{c1} $ as shown in Fig.
1 in \cite{rhh}.

\section{ The evolution of $ C-IC $ in $ h_x $ case }

As motivated in Sec.II-C, we like to investigate possible "generalized " mirror symmetry
around $k_y=\pi/2$. So we apply a shift $k=(0,\pi/2)+q$ to the dispersion Eqn.\ref{YXxen1s} and \ref{abcdhx}
and get
\begin{align}
	&A_q=2+(\cos^2\beta-\sin^2\beta\sin^2\theta)\sin q_y    \nonumber \\
	&B_q=\sin2\beta\sin\theta\cos q_y     \nonumber   \\
	&C_q=\cos q_x       \nonumber  \\
	&D_q=-\sin^2\beta \cos^2\theta\sin q_y
\end{align}
It is easy to see that
the only term which is not mirror symmetric with respect to $q_y=0$
is contained in $A_q$.
($D_q$ has no problem because it is squared in Eqn.\ref{YXxen1s}.)
Making the spectrum mirror symmetric with respect to $q_y=0$ dictates:
\begin{align}
	\cos^2\beta-\sin^2\beta\sin^2\theta=0
\end{align}
   Plugging in the Eqn.\ref{canthx} leads to Eqn.\ref{minihx1}.

Eqn.\ref{minihx1} is obtained demanding that the energy spectrum is symmetric with respect to $k^{0}_y=\pi/2$,
so it guarantees it must be an extreme ( either minimum or maximum ) at $k^{0}_y=\pi/2$ and also the degeneracy condition
$\omega_{k=(0,0)}=\omega_{k=(0,\pi)}$. This explains why Eq.\ref{minihx1}
also contains the C-C$_0$/C-C$_\pi$ boundary Eqn.\ref{minihx3}.

\section{ The evolution of $ C-IC $ in $ h_z $ case }

 Following the procedures in the $ h_x $ case, we will first determine the boundary between $ C-C_0 $ and $ C-C_{\pi} $ by setting
 $ \omega^{-}_{k} (0, 0)=  \omega^{-}_{k} (0, \pi ) $. Using Eqn.\ref{YZxen1s}, we  find it has 4 positive roots $h_1,h_2,h_3,h_4$
 and 4 negative roots. After comparing with numerical results, we find only the two roots $h_2$ and $h_3$ are physical:
\begin{align}
	& h_2=\sqrt{2(3-\cos2\beta)(1-\cos2\beta)}  \nonumber  \\
	& h_3=(3-\cos2\beta)\sqrt{\frac{-2\cos2\beta}{1+\cos2\beta}}
\end{align}
  Setting $h_2=h_3$ leads to $\beta=\beta_*=0.295296\pi$ as shown in Fig.\ref{h3pfine}.
  When $0.25\pi<\beta<\beta_*$, $h=h_3$, when $\beta_*<\beta<\pi/2$, $ h=h_2$.

  Next we determine the constant contour at $k_y^0=\pi/2$, thus we need solve
\begin{align}
	0=\frac{\partial\omega^{-}_k}{\partial k_y}\Big|_{k=(0,\pi/2)}
	\Longrightarrow
	c_8 h^8+c_6 h^6+c_4 h^4+c_2 h^2+c_0=0
\end{align}
 where the coefficients $c_8,c_6,c_4,c_2,c_0$ are functions of $\beta$.
 This equation also has 4 positive roots $ h'_1,h'_2,h'_3,h'_4 $ and 4 negative roots.
 we find only $h'_3$ is a physical solution. Its analytic expression is complicated, so we only show its numerical solution in the
 Fig.\ref{h3pfine}.
 Setting $h_2=h'_3$ leads to $\beta=\beta_0=0.333729\pi$;
 The three lines $h_2, h_3, h'_3 $ and their crossings are drawn in Fig.\ref{h3pfine}.

Since we set $0=\frac{\partial\omega_k}{\partial k_y}\Big|_{k=(0,\pi/2)}$,
the dispersion around $k_y=\pm \pi/2$ changes as shown in Fig.\ref{h3pm}:
\begin{figure}
\includegraphics[width=0.8\linewidth]{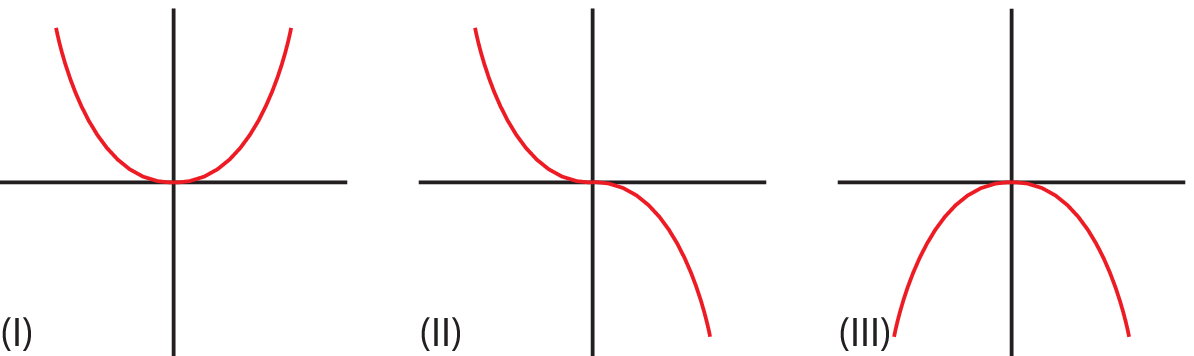}
\includegraphics[width=\linewidth]{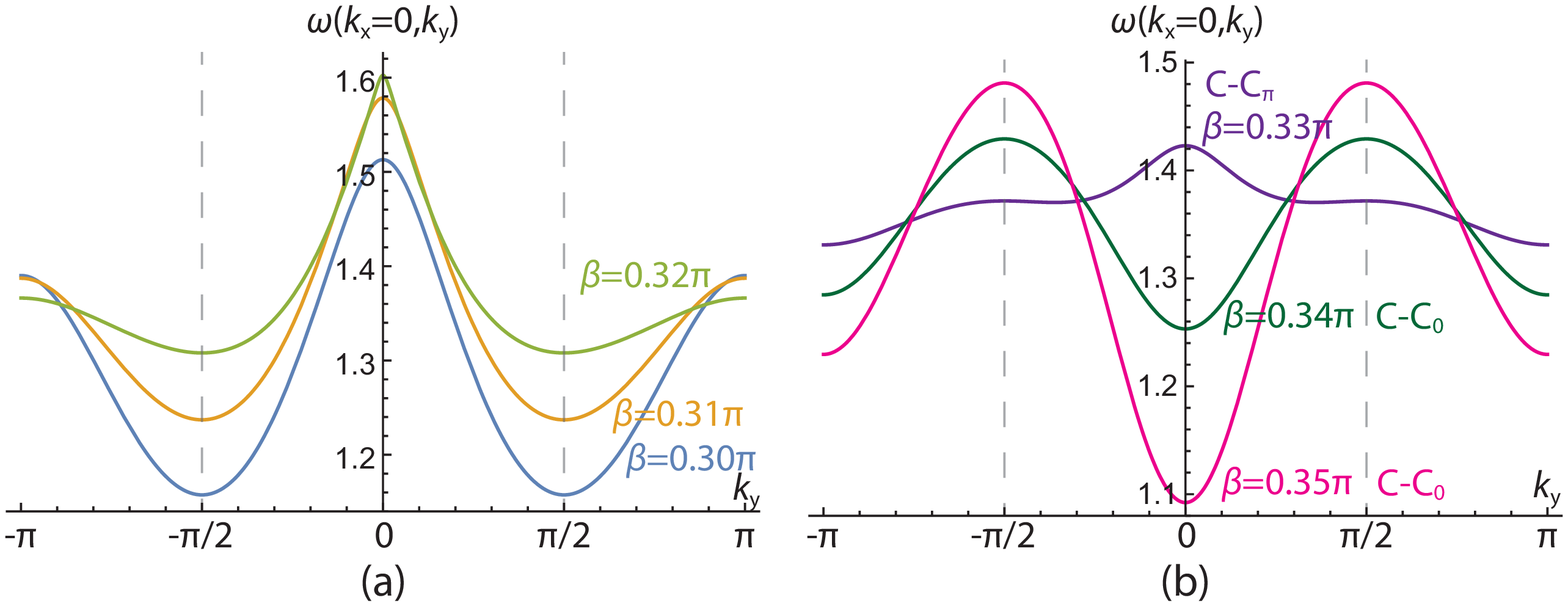}
	\caption{  ( Color online )  (Top)
When $\beta$ increases from $\pi/4$ to $\pi/2$ along $h'_3$,
the contour extreme at $ k^{0}_y= \pm \pi/2 $ goes through (I-II-III) at $\beta_{flat }\approx(0.328\pm0.001)\pi$.
(bottom) The spin-wave dispersion along $h'_3$ for different $\beta$:
	(a)  $0.30\pi$, $0.31\pi$, $0.32\pi$ falling in the range $ \pi/4 < \beta < \beta_{flat} $, $ k^{0}_y= \pm \pi/2 $
is at least a local minimum.
    (b)  $0.33\pi$, $0.34\pi$, $0.35\pi$ falling in the range $ \beta_{flat} < \beta < \pi/2 $.
 $ k^{0}_y= \pm \pi/2 $ becomes at least a local maximum.   }
\label{h3pm}	
\end{figure}

We can summarize the evolution along $h'_3$ line in the following:
Along $h'_3$, when $0.25\pi<\beta<\beta_{\rm flat}$, $(0,\pi/2)$ is a local minimum;
when $\beta_{\rm flat}<\beta<\pi/2$, $(0,\pi/2)$ is a local maximum.

Along $h'_3$, when $0.25\pi<\beta<\beta_{t2}$, $(0,0)$ is a local maximum;
when $\beta_{t2}<\beta<\pi/2$, $(0,0)$ is a local minimum.

Along $h'_3$, when $0.25\pi<\beta<\beta_{t1}$, $(0,\pi)$ is a local maximum;
when $\beta_{t1}<\beta<\pi/2$, $(0,\pi)$ is a local minimum.

The relation between these $\beta$ are
$\beta_{t1}<\beta_{\rm flat}<\beta_{t2}$

If $0.25\pi<\beta<\beta_{t1}$,
$(0,\pi/2)$ is a global minimum;

If $\beta_{t1}<\beta<\beta_{\rm flat}$,
we need compare $(0,\pi/2)$ with $(0,\pi)$;

\qquad
if $\beta_{t1}<\beta<\beta_{tf}$,
$(0,\pi/2)$ is a global minimum;

\qquad
if $\beta_{tf}<\beta<\beta_{\rm flat}$,
$(0,\pi)$ is a global minimum;

If $\beta_{\rm flat}<\beta<\beta_{t2}$,
$(0,\pi)$ is a global minimum;

If $\beta_{t2}<\beta<\pi/2$,
we need compare $(0,\pi)$ with $(0,0)$;

\qquad
if $\beta_{t2}<\beta<\beta_{0}$,
$(0,\pi)$ is a global minimum;

\qquad
if $\beta_{0}<\beta<\pi/2$,
$(0,0)$ is a global minimum;

As summarized in Sec.III-C,
if $0.25\pi<\beta<\beta_{tf}$,
$(0,\pi/2)$ is a global minimum;
if $\beta_{tf}<\beta<\beta_{0}$,
$(0,\pi)$ is a global minimum;
if $\beta_{0}<\beta<\pi/2$,
$(0,0)$ is a global minimum. The final result is shown in Fig.\ref{h3pfine}

\begin{figure}[!htb]
\includegraphics[width=0.48\linewidth]{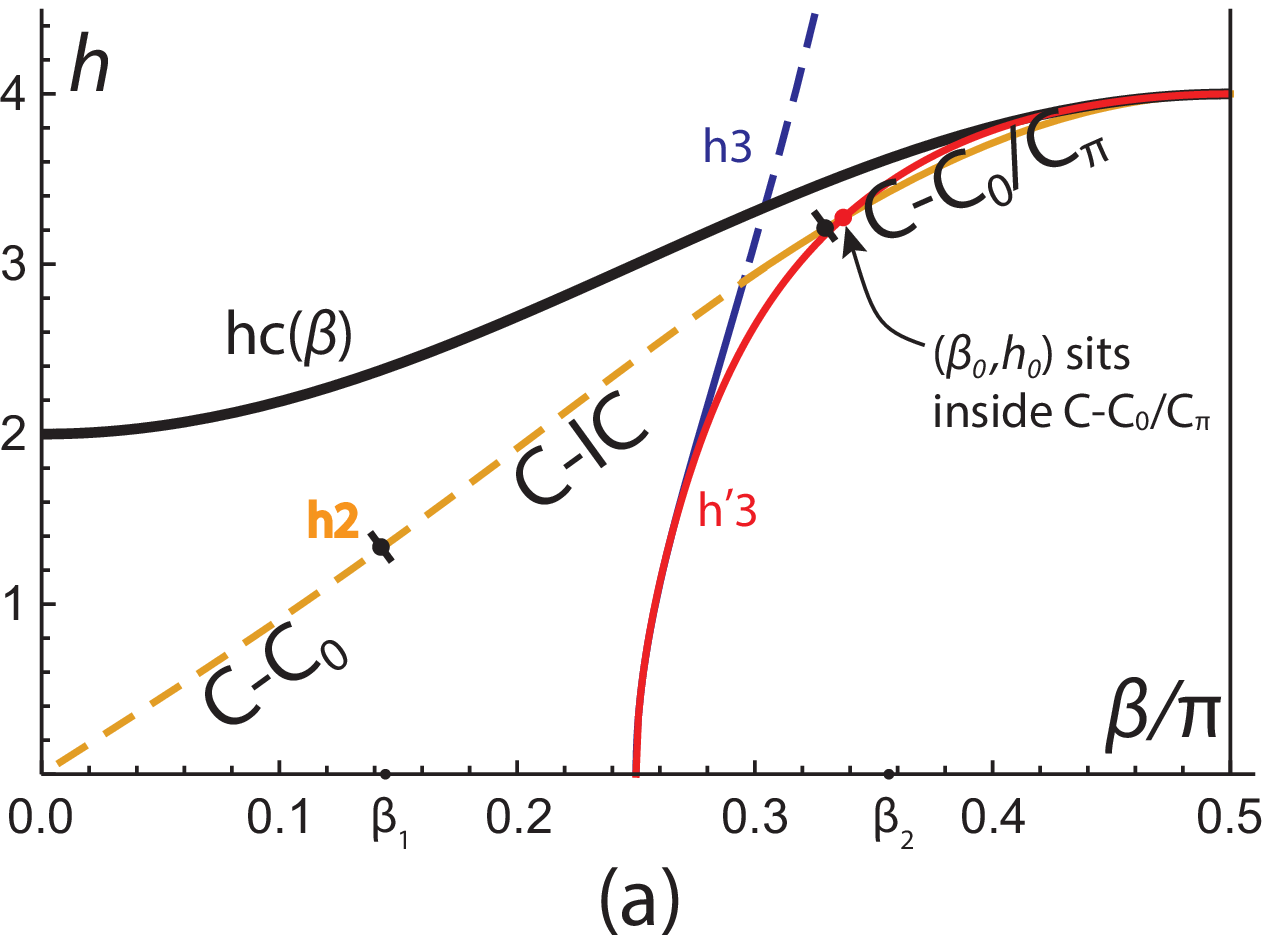}
\includegraphics[width=0.48\linewidth]{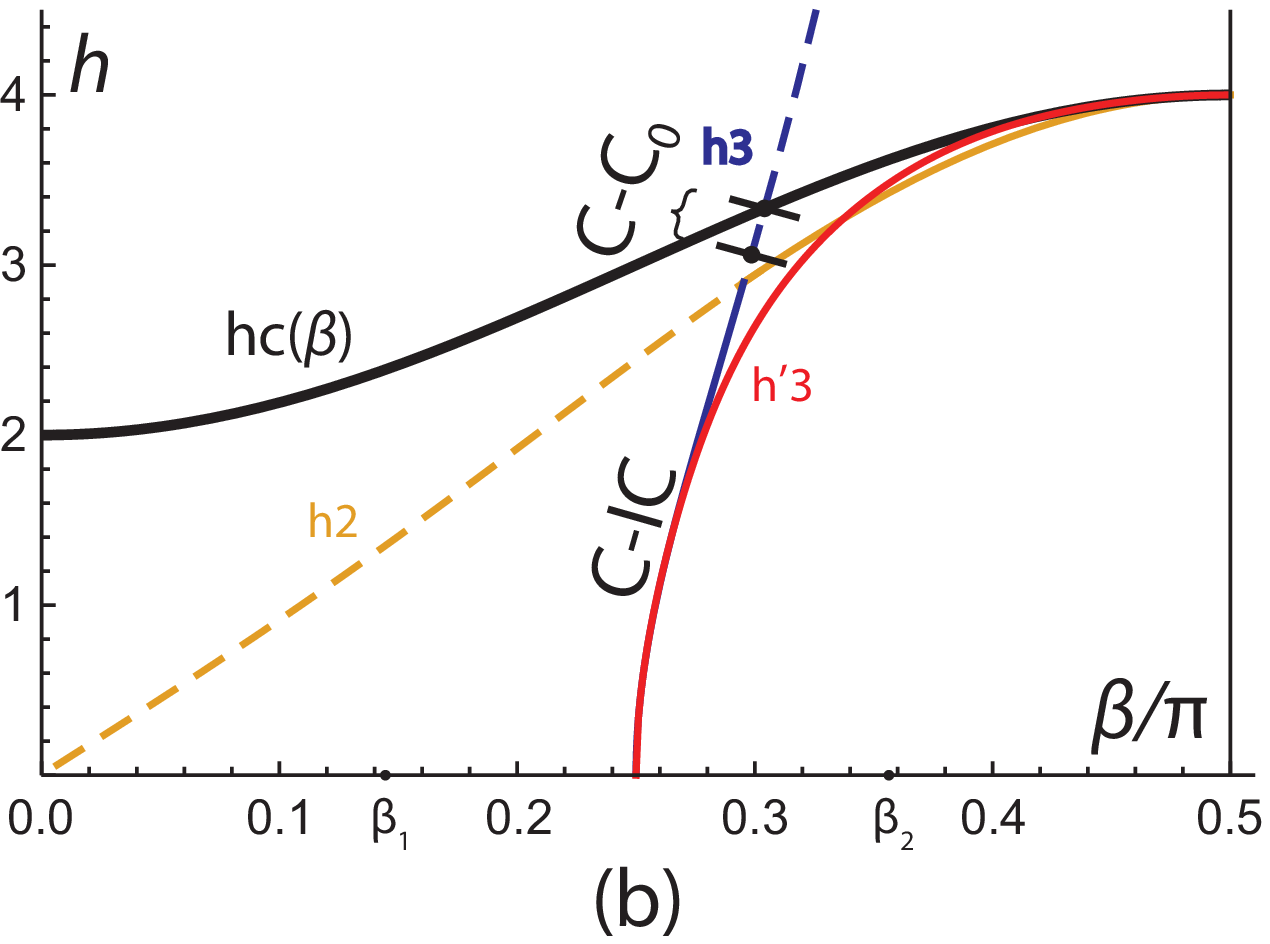}
\caption{  ( Color online ) 
(a) Along $h_2$,
the minimum location of the dispersion is $k=(0,0)$
then become $k=(0,0<k_y^0<\pi)$,
then become either $k=(0,0)$ or $k=(0,\pi)$ which are degenerate along $ h_2 $.
(b) Along $h_3$,
the minimum location of the dispersion is $k=(0,0<k_y^0<\pi)$
then become $k=(0,0)$. }
\label{h2h3}
\end{figure}

 The minimum structure along $ h_2 $ and $ h_3 $ are shown in Fig.\ref{h2h3} and B respectively.

\begin{figure}[!htb]
\includegraphics[width=0.9\linewidth]{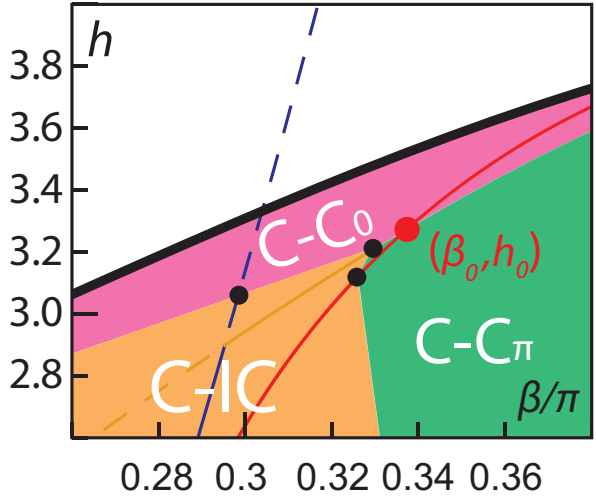}
\caption{   ( Color online ) Fine structure of  C-C$_0$, C-IC, C-C$_\pi$  boundaries around $(\beta_0,h_0)$.
It is reached by connecting those special points  along the three lines $h_2 $ ( solid and dashed brown ), $ h_3 $ (
solid and dashed blue ),$  h'_3 $ ( solid red ) in Fig.\ref{h3pfine} and Fig.\ref{h2h3}. The Thick solid black line is the phase boundary $ h_{cz} $. }
\label{fineh0b0}
\end{figure}

 Combining all the special points along the three lines $h_2, h_3, h'_3 $ in Fig.\ref{h3pfine} and Fig.\ref{h2h3}
 and also $ \beta_1, \beta_2 $ at $ h_x=0 $ lead to Fig.\ref{phasehz}
 and the evolution around $(\beta_0,h_0) $ in Fig.\ref{fineh0b0}.

\begin{thebibliography}{99}

% Magnetisim

\bibitem{scaling} A. V. Chubukov,  S. Sachdev, and  J. Ye,
\textit{Theory of two-dimensional quantum Heisenberg antiferromagnets with a nearly critical ground state},
Phys. Rev. B \textbf{49}, 11919(1994).


% books

\bibitem{aue} A. Auerbach,
\textit{Interacting electrons and quantum magnetism},
(Springer Science \& Business Media, 1994).

\bibitem{wenbook} X.G. Wen,  Quantum Field Theory of Many-body Systems, From the Origin of Sound
to an Origin of Light and Electrons, ( OXFORD UNIVERSITY PRESS, 2004  ),

\bibitem{sachdev} S. Sachdev, \textit{Quantum Phase transitions}, (2nd edition, Cambridge University Press, 2011).

\bibitem{frad} E. Fradkin, Field Theory of condensed matter, ( 2nd edition, Cambridge University Press, 2013).


\bibitem{kitpconf}
Turner, A. M. \& Vishwanath, A.
Beyond Band Insulators: Topology of Semi-metals and Interacting Phases.
Preprint at http://arxiv.org/abs/1301.0330 (2013).

\bibitem{rh}
Fadi Sun, Jinwu Ye, and Wu-Ming Liu,
\textit{Quantum magnetism of spinor bosons in optical lattices with synthetic non-Abelian gauge fields},
Phys. Rev. A \textbf{92}, 043609 (2015).

\bibitem{cwu}
Zi Cai, Xiangfa Zhou, and Congjun Wu, Magnetic phases of bosons with synthetic spin-orbit coupling in optical lattices,
Phys. Rev. A 85, 061605(R), 2012,
For  a review, see, Xiangfa Zhou, Yi Li, Zi Cai and Congjun Wu, Unconventional states of bosons with the
synthetic spin¨Corbit coupling, J. Phys. B: At. Mol. Opt. Phys. 46 (2013) 134001 (16pp).

\bibitem{classdm1} J. Radic¡ä, A. Di Ciolo,  K. Sun, and V. Galitski, Exotic Quantum Spin Models in Spin-Orbit-Coupled Mott Insulators, PRL 109, 085303 (2012)

\bibitem{classdm2} William S. Cole1, Shizhong Zhang, Arun Paramekanti, and Nandini Trivedi,
Bose-Hubbard Models with Synthetic Spin-Orbit Coupling: Mott Insulators, Spin Textures, and Superfluidity, Phys. Rev. Lett. 109, 085302 (20

\bibitem{rhh}
Fadi Sun, Jinwu Ye, Wu-Ming Liu,
\textit{Rotated Heisenberg model in a Zeeman field
and its applications to cold atoms and materials with spin-orbit coupling},
arXiv:1502.05338. under review in Phys. Rev. Lett.


\bibitem{blqh} Longhua Jiang and Jinwu Ye,
Ground state, quasihole and a pair of quasihole wavefunctions in Bi-layer Quantum Hall systems, Phys. Rev. B 74, 245311 (2006).

\bibitem{tri} Jinwu Ye,  Broken symmetry, excitons, gapless modes, and topological excitations in trilayer quantum Hall systems,
Phys. Rev. B 71, 125314 (2005)

\bibitem{tqpt}
Sun, F., Yu, X.-L., Ye, J., Fan, H. \& Liu, W.-M.
Topological Quantum Phase Transition in Synthetic Non-Abelian Gauge Potential: Gauge Invariance and Experimental Detections.
\textit{Sci. Rep.} \textbf{3}, 2119 (2013).


\bibitem{ssdw} Shang-Shun Zhang, Jinwu Ye, Wu-Ming Liu,
 Itinerant magnetic phases and Bosonic Lifshitz transitions in repulsively interacting spin-orbit coupled Fermi gas,
        arXiv:1403.7031, version 5.

\bibitem{pairing} Yi-Xiang Yu, Jinwu Ye, Wu-Ming Liu, Cherence length in attractively interacting Fermi gases with Spin-orbit Couplings,
Phys. Rev. A 90, 053603  (2014).

\bibitem{rafhm} Fadi Sun, Jinwu Ye, Wu-Ming Liu, Hubbard model with Rashba or Dresselhaus spin-orbit coupling and Rotated Anti-ferromagnetic Heisenberg Model,
arXiv:1601.01642


% SO coupling in cold atoms

% cold atom experiments


% SOC experimental proposal

%\bibitem{ZFXu1} Z.-F. Xu, L. You, and M. Ueda, Atomic spin-orbit coupling synthesized with magnetic-field-gradient pulses,   Phys. Rev. A \textbf{87}, 063634 (2013).

%\bibitem{so3d}  B.M. Anderson, {\sl et.al},  Synthetic 3D Spin-Orbit Coupling, Phys. Rev. Lett. 108, 235301 (2012).


%new experiments



\bibitem{expk40} Lianghui Huang, {\sl et.al}, Experimental realization of a two-dimensional synthetic spin-orbit coupling
                 in ultracold Fermi gases, arXiv:1506.02861.

\bibitem{2dsocbec} Zhan Wu, {\sl et.al}, Realization of Two-Dimensional Spin-orbit Coupling for Bose-Einstein Condensates, arXiv:1511.08170.

% Previous works

\bibitem{honey1}
Jiang, L. \& Ye, J.
The mobility of dual vortices in honeycomb, square, triangular, Kagome and dice lattices.
\textit{J. Phys, Condens. Matter} \textbf{18}, 6907 (2006).

\bibitem{dual1}
Ye, J.
Duality, magnetic space group and their applications to quantum phases and phase transitions on bipartite lattices in several experimental systems.
\textit{Nucl. Phys. B} \textbf{805}, 418 (2008).

\bibitem{dual2} Chen, Y. \& Ye, J., Characterizing boson orders in lattices by vortex degree of freedoms.
\textit{ Philos. Mag. } \textbf{92}, 4484-4491 (2012).

\bibitem{dual3}
Ye, J. \& Chen, Y.
Quantum phases, Supersolids and quantum phase transitions of interacting bosons in frustrated lattices.
\textit{Nucl. Phys. B} \textbf{869}, 242 (2013).



% My previous papers


\bibitem{un} Fadi Sun, Jinwu Ye, Wu-Ming Liu, In preparartion.

\bibitem{heat1}
Kinast, J. \textit{et al}.
Heat Capacity of a Strongly Interacting Fermi Gas.
\textit{Science} \textbf{307}, 1296 (2005).

\bibitem{heat2}
Ku, M. J. H. \textit{et al}.
Revealing the Superfluid Lambda Transition in the Universal Thermodynamics of a Unitary Fermi Gas.
\textit{Science} \textbf{335}, 563 (2012).


\bibitem{dosexp} Gemelke, N., Zhang X., Huang C. L., and Chin, C. In situ observation of incompressible Mott-insulating domains in ultracold atomic gases, Nature (London) \textbf{460}, 995 (2009).

% detection method

\bibitem{lightatom1}
Ye, J. \textit{et al}.
Light-scattering detection of quantum phases of ultracold atoms in optical lattices.
\textit{Phys. Rev. A} \textbf{83}, 051604 (2011).

\bibitem{lightatom2}
Ye, J., Zhang, K. Y., Li, Y., Chen, Y. \& Zhang, W. P.
Optical Bragg, atom Bragg and cavity QED detections of quantum phases and excitation spectra of ultracold atoms in bipartite and frustrated optical lattices.
\textit{Ann. Phys.} \textbf{328}, 103 (2013).

\bibitem{hightc1} Jinwu Ye,
Quantum fluctuation generated vortices, dual singular-gauge transformation, and zero-temperature transition from d-wave superconductor to underdoped regime,
Phys. Rev. B 65, 214505 (2002)


\bibitem{hightc2} Jinwu Ye, Thermally Generated Vortices, Gauge Invariance, and Electron Spectral Function in the Pseudogap Regime,
Phys. Rev. Lett. 87, 227003 (2001).


\bibitem{glass1} J. Ye, S. Sachdev and N. Read, A solvable spin glass of quantum rotors,
    Phys. Rev. Lett. 70, 4011 (1993)

\bibitem{glass2} S. Sachdev and J. Ye,  Gapless spin-fluid ground state in a random quantum Heisenberg magnet,
   Phys. Rev. Lett. 70, 3339 (1993)

\bibitem{glass3}  N. Read, S. Sachdev and J. Ye,  Landau theory of quantum spin glasses of rotors and Ising spins,
 Phys.Rev.B,52,  384 (1995

\bibitem{glass4}  M. P. Kennett, C. Chamon and Jinwu Ye, Aging dynamics of quantum spin-glass of rotors,
  Phys. Rev. B 64, 224408 (2001).

\bibitem{pairing2} Yi-Xiang Yu, Jinwu Ye and Ningfang Song, Preprint.

\bibitem{qh}  Jinwu Ye and S. Sachdev, The effects of Coulomb interaction on Quantum Hall critical
      points of systems in a periodic potential,  Phys. Rev. Lett. 80, 5409 (1998).


\bibitem{qhdis} Jinwu Ye, The effects of weak disorders and Coulomb interaction  on Quantum Hall critical points,
Phys. Rev. B60, 8290 (1999).


\end{thebibliography}
\end{document}